\documentclass[amssymb,aps,showpacs,showkeys,amsmath]{revtex4}

\usepackage{graphicx}

\newcommand{\be}{\begin{equation}}
\newcommand{\ee}{\end{equation}}

\newcommand{\bex}{\begin{eqnarray}}
\newcommand{\eex}{\end{eqnarray}}
\begin{document}

\title{A Computational Model for Quantum Measurement}
\author{R. Srikanth}
\email{srik@rri.res.in}
\affiliation{Raman Research Institute, \\
Bangalore- 560~080, Karnataka, India.}


\begin{abstract}
Is the dynamical evolution of physical systems objectively a manifestation of
information processing by the universe? We find that an affirmative answer
has important consequences for the measurement problem. 
In particular, we calculate the amount of quantum information processing
involved in the evolution of physical systems, assuming a finite
degree of fine-graining of Hilbert space. 
This assumption is shown to imply that there is a finite capacity to sustain the immense
entanglement that measurement entails. When this capacity is overwhelmed,
the system's unitary evolution becomes computationally unstable and the system
suffers an information transition (`collapse'). 
Classical behaviour arises from the rapid cycles of unitary evolution and 
information transition. 
	Thus, the fine-graining of Hilbert space determines the location of the 
`Heisenberg cut', the mesoscopic threshold separating the microscopic,
quantum system from the macroscopic, classical environment.
The model can be viewed as a
probablistic complement to decoherence, that completes the measurement process
by turning decohered improper mixtures of states into proper mixtures.
It is shown to provide a natural resolution to the measurement problem and the
basis problem.
\end{abstract}
\maketitle 

\section{Introduction}
Inspite of the tremendous success of standard quantum mechanics, its interpretational
aspects continue to puzzle us, particularly in regard to measurement and
its extrapolation to the macroscopic level \cite{kristan,wz}. Although an overwhelming
number of current studies and applications of quantum mechanics do not depend 
on the philosophical resolutions of these difficulties, they will become
important for understanding future experiments where advancing technology will permit 
one to probe mesoscopic phenomena \cite{ghi98}. Further, they can help 
resolve the measurement problem,  
a subtle but arguably important problem in quantum epistemology 
and foundations. In simple words, it is concerned with understanding how macroscopic 
phenomena, in specific measurement outcomes, are classical even though the underlying
microscopic states exist as quantum mechanical superpositions \cite{hom94}. 
Let a system $S$ to be measured be in the pure state
$
|\psi\rangle = \sum_{i=1}^n a_i |i\rangle$, with $\sum_{i=1}^n |a_i|^2 = 1
$
where $\{|i\rangle\}$ are a complete set of eigenstates of
some observable $\hat{A}$ that can be measured by a measuring apparatus $M$. 
Formally, measurement is represented by the action of
the projection operator $\pi_i = |i\rangle \langle i|$ on $|\psi\rangle$,
with probablity $|a_i|^2$, 
and more generally, by the action of arbitrary measurement operators \cite{nc,pre00}.
Under such action, the system $S$ is 
said to undergo an irreversible `reduction of the state vector',
`collapse of the wavefunction' or `quantum \mbox{leap/jump}' to an eigenstate.
In contrast, Schr\"odinger evolution is given by the deterministic
action of a unitary operator. % \cite{zeho}.
This difference in the actions of unitary evolution and measurement is the
simplest manifestation of the measurement problem.

To be precise, we need to take into consideration the measuring apparatus.
Following von Neumann \cite{von71}, we visualize the measurement as beginning with $S$ 
interacting with $M$
in the `ready' state $|R\rangle$. During this ``pre-measurement" phase, the
interaction Hamiltonian entangles $S$ with $M$. 
Assuming the apparatus can be characterized
by a single degree of freedom, represented by the states
$\{|\xi_i\rangle\}$ that span the pointer basis, one obtains the state
\be
\label{envent}
|\psi\rangle|R\rangle = \sum_i a_i|i\rangle|R\rangle \stackrel{\hat{H}}{\longrightarrow} 
|\Psi\rangle = \sum_i a_i|i\rangle|\xi_i\rangle,
\ee
whereby every state $|i\rangle$ gets correlated with a definite macroscopic apparatus 
state $|\xi_i\rangle$. Eq. (\ref{envent}) implies a
superposition of {\em macroscopic} configurations of $M$, contrary to our
everyday experience of determinateness of macroscopic objects. 
Expressed verbally, how/when do the
logical {\tt AND}'s (the +'s in the summation over $i$) in Eq. (\ref{envent}) becomes
{\tt OR}'s? This is the measurement problem. 
As recently demonstrated in Ref. \cite{bas00}, such a projective operation
represents a break-down in the
principle of linear superposition for state vectors, irrespective of the details of
the measuring apparatus and the measurement process.  

The oldest attempt to make sense of this dual dynamics in quantum mechanics is the
Copenhagen interpretation, due mostly to Bohr and Heisenberg \cite{kristan}. It was
not a single interpretation, but rather a collection of somewhat different viewpoints, 
the main
thrust being that physical systems evolve as quantum objects, whereas measuring
apparatuses and their outcomes are to be represented classically:
if a quantity ${\cal Q}$ is measured in system $S$ at time $t$ then 
${\cal Q}$ has a particular value in $S$ at time $t$.
A difficulty with the Copenhagen interpretation appears to be that
it divides the physical world into the microscopic, quantum realm of physical systems, 
governed by the Schr\"odinger equation and linear
superpositions, and the macrocopic, classical realm of 
measuring apparatuses, governed by Newtonian dynamics.  
However, this pragmatic and admirably minimalist approach 
does not clarify exactly where this ``Heisenberg 
cut" between the quantum system and the classical  
measuring apparatus or observer lies, and moreover how and also why its existence 
does not weaken the premise that all phenomena should be describable as
quantum mechanical. % \cite{bel90}. 
As a criticism of this dual unitary-and-projection dynamics, Schr\"odinger proposed 
his well known cat paradox \cite{schr,hom94}. 

To account for measurements, since $M$ in Eq.
(\ref{envent}) does not admit a classical description, 
von Neumann \cite{von71} introduces a second apparatus $M^{\prime}$
which observes and collapses $M$, and possibly a third apparatus $M^{\prime\prime}$, 
and so on until
there is a final measurement, which has a definite result and is not described by 
quantum mechanics, except in the statistical sense.
The Bassi-Ghirardi theorem \cite{bas00} implies that such a final measurement which terminates 
this chain of {\em efficient causes} is inevitable. This line of argument suggests that the
``cut" between quantum and classical worlds can be postponed to various depths of
measurement, but not avoided.

An important role in the measurement process is played by 
(environment induced) decoherence \cite{zur91}, the
practically irreversible quenching of certain off-diagonal elements 
from the density matrix of a bounded but open system \cite{joo85,deco1,zur91,decoz,
per000a,kok01}. It can be shown to
arise naturally as a result of a system's unavoidable interaction and entanglement
with its environment according to the Schr\"odinger equation for a global system
under certain initial conditions. 
Given $\{|e_i\rangle\}$ as states of the environment $Q$ correlated with $M$'s pointer
states $\{|\xi_i\rangle\}$, after some time, process (\ref{envent}) leads to
$|\Psi\rangle|0\rangle \longrightarrow \sum_i
a_i|i\rangle|\xi_i\rangle|e_i\rangle$. The resulting density operator is:
\be
\label{dens}
\rho_{SMQ} = \sum_{i,j} a_ia^*_j|i\rangle|\xi_i\rangle|e_i\rangle
\langle j|\langle \xi_j|\langle e_j|,
\ee
Detailed analysis of the problem shows that the states of $SM$ rapidly decohere,
i.e., states $\{|e_i\rangle\}$ can be
treated as if they are orthogonal \cite{per000a}. Next, tracing over the (unknown)
environmental degrees of freedom, 
and letting $\Pi_i \equiv |\xi_i\rangle \langle \xi_i|$ on $M$,  we obtain:
\begin{subequations}
\label{projex}
\bex
\rho_{SM} \rightarrow \hat{\rho}_{SM} &=& {\rm Tr}_Q(\rho_{SMQ}) 
 = \sum_{i} |a_i|^2|i\rangle\langle i|\otimes|\xi_i\rangle\langle \xi_i| \label{projext}\\
&=& \sum_i \Pi_i \rho_{SM} \Pi_i. \label{projexp}
\eex 
\end{subequations}
Thus, the result (\ref{projexp}) of a {\em non-selective} (projective) measurement on 
$\rho_{SM}$, is the same as the reduced density operator in Eq. (\ref{projext}).
Hence, decoherence indeed resolves 
the measurement problem vis-a-vis non-selective measurements.

We note that Eq. (\ref{projex}) conflates two different meanings of
the reduced density operator: on the one hand, 
the apparent ensemble (\ref{projext}) resulting from
tracing over $\{|e_i\rangle\}$, and on the other, a genuine classical mixture 
of pure states in (\ref{projexp}). Following d'Espagnat
\cite{esp}, we will call the former an ``improper mixture", and the latter
a ``proper mixture". This distinction is not without some controversy, but
will be found convenient for discussing different models addressing 
the measurement problem. In particular, decoherence in standard quantum mechanics
by itself would imply that all mixtures are ``improper".
Whereas this can effectively account for non-selective measurements, it encounters  
some difficulty in accounting for 
individual {\em selective} measurements, which are conditioned on a read-out:
namely,
why all diagonal elements of the reduced density matrix except one vanish
in a given single measurement episode with a known outcome?
In a given measurement whose outcome has been read-out and is 
{\em known} (which is what measurement is all
about), the measured system $S$ is formally represented by a pure state. 
Since the reduced density operator for an entangled
system is necessarily mixed, a pure state cannot be entangled
with any other system, known or unknown. 
This line of reasoning suggests that a procedure like Eq. (\ref{projex}), although
able to account for non-selective measurements, is unable
to account for selective measurements.
A more detailed discussion of the problem is considered by Adler \cite{adl01}.

One other possible solution to this situation is to 
invoke the relative-state \cite{everett} or 
many-worlds \cite{dewitt} or similar interpretation of quantum measurement. 
The relative state interpretation \cite{everett} of quantum measurement
rejects the projection postulate,
instead postulating that each entangled $SM$+observer correlated
branch in Eq. (\ref{envent}) is realized in a different version of reality.
Applying it to the decohered system, one argues
that a given state $|i\rangle|\xi_i\rangle|e_i\rangle$ of $SMQ$
is pure relative to the observer in a given
version of the universe, but mixed across the different versions in the
multiverse. Thus the mixture $\hat{\rho}_{SM}$ resulting from decoherence
is to be understood in the special sense as existing due to entanglement
across the multiverse, to be conceptually contrasted from a
conventional mixture of pure states existing in the given same version of the 
universe. 
The main difficulty with the Everett theory is that it is
not clear how it is supposed to account for the apparent determinateness 
of measurement outcomes \cite{barstan},
while accounting for the statistical nature of measurement outcomes \cite{multiverse}. 
Attempts to reconstruct the Everett theory to address these problems have 
led to detailed, interesting 
formulations of quantum mechanics as the many-worlds \cite{dewitt},
many-minds \cite{alblow}, consistent or decoherent histories
\cite{gri86,om88,gelhar}, and relative-facts 
theories \cite{rf}. Some difficulties in their interpretation have been discussed 
\cite{obj}.
Other novel solutions proposed include % the transactional interpretation
% \cite{cramer}, an interpretation invoking objective fuzziness \cite{mohr}, 
the restricted path-integral method \cite{men02} and a Grover search-based method
\cite{man02}. 

The Bohmian model \cite{boh62} offers a
different but related mathematical formalism from standard quantum mechanics, based
on a nonlocal deterministic picture, in which there is no discontinuity between
ordinary evolution and measurement. In the collapse or 
dynamical program,
\cite{ghi,ghistan} and the gravitational model by Penrose \cite{pen96}, 
the projection postulate is derived as a consequence of additional physics. In the
former, this is achieved via a small stochastic, nonlinear term added to the 
dynamical equation of the standard theory; in the latter, via an
energy uncertainty due to 
gravitational field superpositions due to states that are
spatially apart. 
 
The present work aims to resolve the measurement problem 
within the decoherence scenario, by examining the information
theoretic foundations of physics.
The model of quantum measurement proposed here
is primarily based on the premise that the content and evolution of physical
phenomena are well characterized as an abstract registration and processing of quantum
information, not all of which is physically accessible, on account of limitations 
imposed by quantum uncertainty; and that,
from a certain point of view, such an approach can be as useful to describe
physical phenomena as the conventional mathematical language of differential
equations. We begin by calculating bounds on the quantum information 
processing involved in the evolution of quantum states in a Hilbert space with
finite fine-graining.
As the next step, we study the implications of these bounds exceeding finite
memory and computational resources available to physical systems, as determined
by the system's energy and Hilbert space resolution.
We suggest that the measurement problem finds a 
natural resolution as a consequence of 
the finiteness of quantum computational parameters characterizing physical dynamics.
Thereby, the quantum to classical transition emerges as a
manifestation of the underlying computational structure of physical systems.

It is of interest to note that in a number of recent 
works \cite{llo00,llo02,siv00,ng,pat02}, bounds on the amount of information
processing that can be performed by physical systems have been obtained. In
particular, Lloyd \cite{llo00} employs the Margolus-Levitin theorem \cite{marlev98}
to show that the number of elementary logical operations per second
(``op/s") that a physical system can perform is limited by its energy, while the amount
of information that it can register bounded by its maximum entropy \cite{lan88,bek}.
Sivaram \cite{siv00} studies the implications of model independent features of
quantum gravity for everyday physics problems like field strengths, temperature,
acceleration, particle energies and the performance of clock and computer
performance. Ng \cite{ng} has argued that the product of speed and clock rate in a
computer is bounded above by $t_P^{-2}$, where $t_P = \sqrt{\hbar G/c^5} =
5.4 \times 10^{-44}$ sec is the Planck time. Pati et al. \cite{pat02} show that
the total number of operations performed by a computer is determined by the type
of interactions present in it, a fact useful in designing and building a quantum
computer. 

To be precise, the information dealt with in the preceding works is the 
accessible and useful information. Because of quantum uncertainty, this quantity
is much smaller than the information needed to support or describe physical states
of many-particle systems.
An attempt to access this larger information in a quantum system in general
`collapses' the state, and only a small fraction, bounded by the von Neumann information
\cite{nc,pre00}, can actually be extracted for useful information tasks.

The bounds in the 
numbers of op/s and bits we obtain can be interpreted in three distinct ways: (i) They
give upper bounds to the amount of quantum computation that physical systems
perform; (ii) they give lower bounds to the number of op/s and bits required to
simulate physical systems; (iii) If one chooses to regard nature as
performing computations, these numbers give the op/s and bits in that computation.
In particular, this last idea is used to motivate the 
`reduction' of the state vector under measurement, and the
emergence of classicality at the macroscopic level.
In Sections \ref{mem} and \ref{spee}, we first 
obtain bounds on the amount of quantum information processing needed to
support physical systems, and the implications of entanglement on these bounds
in Section \ref{entbound}. In Sections \ref{compu} and \ref{mez}, we present a 
computational model of measurement in which the observed classicality of 
measurement outcomes is related to parameters characterizing the information processing
capacity of physical systems. This is specifically applied to projective
measurements in Section \ref{vonn}. The related issue of emergence of classicality in
the macroscopic world and that of macroscopic quantum interference effects 
are dealt with in Section \ref{eme}. 

We feel obliged to apologize in advance for the 
many instances of usage of computer theoretic descriptions of purely
physical processes, sometimes accompanied with only qualitative 
justification, that the reader might encounter in this article. Yet, we believe
our approach is justified, as the present model
is basically about applying computer/information theoretic intuition to physical
processes. Expositions of issues at a qualitative level are
sometimes unavoidable, since, as we saw, some subtle
issues surrounding the measurement problem
hinge basically on physical interpretation. 

\section{Memory allocation depends on Hilbert space resolution}\label{mem}

The amount of classical information required to specify even the simplest quantum 
system, a qubit, is obviously infinite. More generally, we can say that the amount of 
information (in bits) required to prepare or specify a
quantum state vector-- the preparation information or {\em state information}--
depends on the chosen
fine-graining of the Hilbert space ${\cal H}$ in which the quantum 
state vector resides. 
The chosen precision at which we want to resolve a system's ${\cal H}$ is
quantified by $\mu$, the number of bits per probability amplitude needed to represent a 
state vector. Of these, $\mu/2$ bits specify the real part, 
and the remaining $\mu/2$ bits the imaginary part.

In the geometric formulation of
the Hilbert space \cite{woo,bra94}, the minimum seperation between 
two microstates (i.e, state vectors) is given by $\phi = 2^{-\mu/2}$,
where $\phi$ is the smallest resolvable Hilbert space angle, a measure of 
distance given by
the Fubini-Study metric, a Riemannian metric defined on projective Hilbert
space \cite{sch94}. Eg., for a qubit specified at $\mu = 2$ bytes precision,
state vector resolution is a ``sphere" of size 
$2^{-\mu}\pi \simeq 4.8 \times 10^{-5}$ radians.  For a uniform
ensemble of all resolvable state vectors $|j\rangle$, 
each with probability $p_j = 2^{-\mu}$, on average and $D \equiv {\rm dim}{\cal H}$, the
quantity:
\be
\label{sv}
H(\tilde{p}) \equiv - \sum_{j=1}^{D-1} p_j \log p_j = (D - 1)\mu ~{\rm bits}
\ee 
is the state information required to construct an arbitrary $|j\rangle$.
In Eq. (\ref{sv}), $\log()$ refers to base-2 logarithm.
In general,  $H(\tilde{p}) \ge S(\rho )$, the von Neumann entropy \cite{nc,pre00}, 
because $S(\rho)$ measures entropy with basis states as the
statistical alternatives, whereas $H(\tilde{p})$ measures entropy with all
resolvable state vectors as the alternatives. Therefore, $H(\tilde{p}) =
S(\rho )$ when the ensemble consists only of orthogonal states.

For an isolated system of $D$ (finite) 
dimensions, the number of bits sufficient to specify 
the state fully at $\mu$-bit precision is given by
Eq. (\ref{sv}), in view of the normalization condition. 
However, for our purpose, it is preferable to write down ${\cal M}$, 
the memory required to register the state information at any time, as 
\be
\label{qmem}
{\cal M} = D\mu ~{\rm bits}.
\ee
This turns out to be the more efficient specification taking into consideration the
time evolution of a quantum state, discussed in the next two Sections.
The additional computational complexity for evaluating an amplitude via the
normalization condition at each step goes as $\sim O(D^n)\mu$, which increases 
exponentially for a system of $n$ objects, 
whereas unitary evolution of the amplitude requires only $\sim O(\mu)$ steps.
Thus the exponential increase in computational complexity that would result 
with the saving in memory outweighs the linear increase in memory usage. 
The basis in which the $D$ amplitudes are taken to be specified is some fixed, 
arbitrary reference basis. Later we will find that because of 
the finiteness of $\mu$ and decoherence, a preferred basis can emerge at the
macroscopic level.

Although we have derived ${\cal M}$ as a Shannon information for microstates in Hilbert 
space, in fact it is better viewed as absolute information in the sense of 
Kolmogorov-Chaitin complexity \cite{chaikol}. 
Based on a modern notion of randomness
dealing with the quantity of incompressible information in individual objects, 
their algorithmic information is a form of
absolute information contained in them. 
It is a measure of pointwise randomness rather than average
randomness produced by a source, which is the
primary concern of Shannon
information and its quantum generalization, von Neumann information.
Thus, algorithmic information characterizes a system in a way impossible to
classical probability theory (a branch of measure theory
satisfying the Kolmogorov axioms). 
It employs the notion of the length of the shortest effective description of an
individual object, which is its Kolmogorov-Chaitin complexity, 
to quantify the randomness of individual objects in an objective and absolute manner. 
We characterize ${\cal M}$ as the algorithmic information of the state vector.
It is absolute because it depends only on $D$ and $\mu$, and not on the particular 
state or mixedness of the state. Thus, the pure states $|0\rangle$,
$\cos\theta|0\rangle + \sin\theta|1\rangle$ and the maximally mixed state 
$(1/2)(|0\rangle\langle0| + |1\rangle\langle1|)$ all require the same ${\cal M}$:
namely, $2\mu$. This basically reflects the idea that from the viewpoint of nature,
viewed as the underlying information processor, all states are pure.
It is characteristic of quantum mechanics that the maximum information
accessible is far smaller than ${\cal M}$, restricted in fact by the Holevo bound
\cite{nc,pre00}.

For `high fidelity' evolution,
the precision parameter $\mu$ should be sufficiently large in relation to the
dimension of ${\cal H}$. In particular, consider a uniform superposition in some
basis. Each eigenstate has a probability $D^{-1}$ of being found. 
Each amplitude, and hence probability, is specified with accuracy $\mu/2$.
According to the {\em computational completeness principle}, the fine-graining
should be large enough to specify a uniform superposition, i.e.,
$D^{-1} > 2^{-\mu/2}$, or $2\log D < \mu$. In practice,
for reasonably accurate state representation, we require $\mu \gg 2\log D$, 
but we will use the above weaker condition as a thumb rule.

We define total state information of an elementary particle
as the algorithmic information required to
specify the complex wavefunction $\psi(x,t,\Gamma)$ of a physical system over a lattice
of Planck-sized cells filling all configuration space.
% This assumes that quantifying ${\cal M}$ is ultimately linked to 
% the appropriate theory of quantum gravity,
% and thereby to natural barriers like Planck length.
Here $\Gamma$ represents the
internal degrees of freedom. Note that the choice of position basis is quite
arbitrary, and one could as well evaluate ${\cal M}$ using the momentum basis wavefunction
$\psi(p,t,\Gamma)$. 
Let us consider the state information for a free electron, treated as an
elementary spinor particle. In position basis, the spatial part of its state is written:
$|\psi\rangle = \sum_{x=1}^{\cal N}f_x|x\rangle$ with 
$\sum_{x=1}^{\cal N}|f_x|^2 = 1$,
where $|x\rangle$ is the position eigenstate for a particle being localized at
Planck cell $x$ in the universe, 
${\cal N}$ is the number of $l_P^3$ sized cells given by 
$(4\pi/3)(ct_U/l_P)^3 \approx 2.8 \times 10^{181}\mu$, where $t_U = 10^{17}$ s
is the lifetime of the universe. Taking into consideration the spin part, we have
$D = 2{\cal N} \approx 10^{182}$.
Therefore ${\cal M} = D\mu = 10^{182}\mu 
\approx 2^{600}\mu$ bits. Therefore, according to the computational completeness
principle, we must have
$\mu > 600$ bits. It is interesting that this criterion requires us to relate a
cosmological number with the property of an electron's state representation,
somewhat reminiscent of Dirac's large number hypothesis \cite{llo02}. 

Though ${\cal M}$ is a measure of the complexity of a system, it is different
from the number of states $\exp(S/k_B)$, estimated from thermodynamic entropy $S$
or density of states estimated quantum mechanically. For ${\cal M}$ not only
quantifies the number of states available to a system, but also takes into the
consideration the spatial information needed to specify those states. Thus, as a rule
${\cal M} \ge \exp(S/k_B)$.
In view of this, one might ask whether a measure of complexity based on a
suitable conventional definition of entropy
can be considered as a compressed, and hence more economic, form of the
${\cal M}$. The answer is two-fold: in most cases where density of states is
evaluated, we often assume simplifying symmetries and time-independence. A given
state (eg., the $S^1$ orbital in the H atom) can be deformed in a large number
of ways as the atom interacts with its environment, none of which can be
accounted for in a counting system that numbers only states. Moreover, from a computer
theoretic viewpoint, which is relevant to the present model, to
compute the dynamics at each time slice from such compressed information
requires that the information be uncompressed at each time slice to compute
the further dynamics. As explained earlier, this invariably implies an
exponential increase in computational complexity. For these reasons, state
information as defined above, is suitable for our purpose.

The state information may be contrasted
from the amount of information that a physical system (a) can in principle
register, and (b) the quantity of information
that it is possible in practice to store and retrieve from a physical system,
taking into consideration noise temperature, cost of transporting a bit, etc.
\cite{von,lan88,llo00}. As computed earlier, such a particle
requires ${\cal M} \ge 2{\cal N}\mu = 2^{600}\mu$ bits of state information. 
As to (a), it is the von Neumann entropy
bounded above by $\log D$, where $D$ is the dimension of 
the relevant Hilbert space. A free electron can in principle 
encode for $\log(2{\cal N}) \approx 600$ bits. 
As to (b), it is
$I = S/k_B\ln(2) = 3/2\ln(2)$ bits, where $S = k_BTd/2$ is entropy, and
$k_B = 1.38 \times 10^{-23}$ J/K, $T$, $d$ are, respectively, 
Boltzmann constant, temperature of system, and degrees of
freedom available to it. Eg., the maximum information registrable in a particle's
kinetic energy is 3/2 bits. These two informational quantities are far smaller 
than ${\cal M}$, which is not entirely accessible because of quantum uncertainty
and quantum no-cloning \cite{woo82}. 

It is instructive to compare ${\cal M}$ for a free elementary particle
with the information registered by
conventional computers. Estimating the total number of computers to be $\approx
10^9$, each registering $10^{12}$ bits, yields $10^{21}$ bits, which is vastly less
than the $\approx 10^{182}\mu$ bits of state information required to completely specify 
a single free elementary particle.
Indeed, this even exceeds the usable 
number of bits available in the entire universe, which is about $10^{120}$ bits,
including the gravitational degrees of freedom \cite{llo02}.

\section{Computational speed depends on energy and resolution}\label{spee}

The maximum number of operations that can be performed by a physical system is
proportional to its energy \cite{llo00,marlev98}. This result follows
essentially from the observation that the speed with which a quantum system evolves is
determined by its Hamiltonian, and hence its average energy.
As an example, consider the evolution of a qubit with logical states $|0\rangle$
and $|1\rangle$ on which we perform the NOT operation. To flip the qubit one can
apply a potential $\hat{H}_0 = E_0|E_0\rangle\langle E_0| + E_1|E_1\rangle\langle E_1|$
with energy eigenstates $|E_0\rangle = (1/\sqrt{2})(|0\rangle + |1\rangle)$ and 
$|E_1\rangle = (1/\sqrt{2})(|0\rangle - |1\rangle)$. Because 
$|0\rangle = (1/\sqrt{2})(|E_0\rangle + |E_1\rangle)$ and 
$|1\rangle = (1/\sqrt{2})(|E_0\rangle - |E_1\rangle)$, each logical state has an
energy spread $\Delta E = (E_1 - E_0)/2$. Under application of the
potential, the system prepared in state $|0\rangle$ for a small time $t$ becomes:
\be
\label{evolution}
|\Psi(t)\rangle = \frac{1}{\sqrt{2}}\left(|E_0\rangle + e^{i2\Delta
Et/\hbar}|E_1\rangle\right).
\ee
It follows from Eq. (\ref{evolution}) that the time taken to flip the
qubit to $|1\rangle$ is given by $\pi\hbar/2\Delta E$.
Similarly, it is easy to verify that after time $\pi\hbar/2\Delta E$, the qubit evolves 
so that $|1\rangle \longrightarrow |0\rangle$. 
Now the states $|0\rangle$ and $|1\rangle$
are mutually orthogonal. The average energy $E$ for both states
$\langle 0|\hat{H}_0|0\rangle = \langle 1|\hat{H}_0|1\rangle = (E_0 + E_1)/2 = E_0 + \Delta E$.
Thus, both the average energy and spread in energy limit the computational speed of
a physical computer. By embedding a Toffoli or Controlled-Controlled-NOT 
gate in a quantum context, one can
similar show that the AND and FANOUT gates, and thus a universal set of gates for
classical computation, can be performed at a rate $f = 2E/\pi\hbar$.
Therefore, the rate at which a quantum computer
processes classical information is $f = 2E/\pi\hbar$,
where we have set $E_0 = 0$, so that $\Delta E = E$ \cite{marlev98}. 
Here, by `classical' we mean information recorded in orthogonal states. In
principle, such information can be encoded, processed and read-out with complete
certainty.

The evolution of state information is clearly much faster. Consider a
$D$-dimensional system, driven by the Hamiltonian
$\sum_{i=1}^DE_i|E_i\rangle\langle E_i|$. Along any given `direction' in Hilbert
space, the minimum distance is $\phi=2^{-\mu/2}$. The minimum time to traverse
from one microstate to the neighboring is $2^{-\mu/2}\hbar/E_i$. Thus, the $i$th
amplitude evolves at the speed $f(i)=2^{\mu/2}E_i/\hbar$. 
Therefore, the speed in op/s at which amplitudes are in effect
computed as a physical system evolves in real time is:
\be
\label{qfreq}
f_q = \frac{2^{\mu/2}DE}{\hbar}~~{\rm op/s},
\ee
where average energy $E=\sum_i E_i/D$.
We can view $f_q$ as the bit rate at which $D$
amplitudes of a system with energy $E$ are being evolved at $\mu$-bit precision.
For a two-state system $f_q = 2^{\mu/2}\pi f$. The implicit quantum information 
processing proceeds
exponentially faster in $\mu$ than usable classical information processing.

Applying this result to an electron with energy $E = mc^2
\approx 8.2 \times 10^{-14} J$ shows that an electron can
in principle process classical information at the rate $f = 8.6 \times 10^{20}$.
If the electron is treated as a qubit, with two dimensions, and amplitudes are
specified at double precision, i.e., $\mu = 64$ bits, then the speed of computation
implied by an isolated electron's evolution is $f_q = 2^{32}f =
3 \times 10^{30}$ elementary logical operations per second. More realistically,
if the electron is treated as an elementary spinor particle, 
the speed of computation 
is $f_q = 10^{\mu/2}10^{182}(8.2 \times 10^{-14})/(6.1 \times 10^{-34})
\approx 10^{202}2^{\mu/2}$ op/s.  Assuming that amplitudes
are specified with the minimal 600 bit-precision, 
$f_q = 10^{202}2^{300} \approx 10^{292}$ op/s.

Interestingly, the computational speed of all conventional computers combined is
far lesser than the number of logical
operations performed in the quantum evolution of an single electron. With about
$\approx 10^9$ computers operating at a clock rate of $\approx 10^9$ Hz
performing $\approx 10^5$ elementary logical operations per clock cycle, all the
human-made computers in the world are operating at no more than $10^{23}$ op/s.
This is only 3 orders larger than the rate at which classical logical operations
are preformed by
a single electron, but about 270 orders smaller than processing rate corresponding to
the evolution of an electron at the minimal, 600-bit resolution of ${\cal H}$. 
In this and the previous Section,
we saw that energy and Hilbert space resolution limit the computational speed and
memory support required as a quantum system evolves. 
In the following Section, we will find that these 
needs are exponentially augmented in the presence of entanglement.

\section{Entanglement augments speed and memory requirement}\label{entbound}

In practice, few physical things in the universe are truely isolated. Systems are
constantly interacting and, as a result becoming entangled. 
Consider the interaction of $n$ $D$-dimensional objects, each characterized by
Hilbert space ${\cal H}_D$. When isolated, each has a memory
requirement ${\cal M} = D\mu$ bits. If they do
not interact, and thus remain separable, 
their total memory requirement is simply $nD\mu$ bits. But if they do, they
will in general become entangled, and the combined system is characterized by
the $D^n$-dimensional Hilbert space ${\cal H}_D^{\otimes n}$, so
that the general state is now described by the exponentially larger
$D^n$ amplitudes. This is the familiar
consequence of the tensor product character of Hilbert spaces \cite{nc,pre00},
and reflects the fact that multiparticle systems, having more permutations of
configuration, are exponentially more complex than isolated systems. 
The memory requirement for the system, composed of elementary
objects, is given by
\be
\label{qnmem}
{\cal M} = \mu D^n \equiv \mu{\cal D} ~~{\rm bits}.
\ee 
If the dimensions of the interacting elementay objects is $D_1, D_2,\cdots$, 
then ${\cal D} = \Pi_iD_i$, and ${\cal M} = {\cal D}\mu$. 
On the other hand, if the objects are
non-interacting, and thus remain separable, then ${\cal M} = (\sum_iD_i)\mu$. 

One might very loosely regard ${\cal M}$ as a measure of 
entanglement in the system in the limited sense that given a $n$-partite state,
it is larger for an entangled state than for a separable state. 
However, it is an absolute quantity, 
depending only on $n, D$ and $\mu$, and not on the particular
state or mixedness of the
state. Entangled states with different conventional
quantifications of entanglement can have same the ${\cal M}$, whilst 
states quantified with
the same entanglement conventionally can have different ${\cal M}$.
For example, the states $(1/\sqrt{2})(|00\rangle +
|11\rangle)$ and $\cos\theta_1\cos\theta_2|00\rangle + 
		  \cos\theta_1\sin\theta_2|01\rangle + 
		  \sin\theta_1\cos\theta_3|10\rangle + 
		  \sin\theta_1\sin\theta_3|11\rangle$ 
($\theta_1 = \pi/3,\theta_2=-\theta_3=\pi/4$)
are quantified to have two different amounts of (entropy of) entanglement \cite{bruss}
(1 bit and 0.81 bits, respectively), but their state information is stored using the 
same amount of memory: ${\cal M} = 4\mu$ bits. This reflects the fact
that when $n$ objects are entangled, $\Pi_iD_i$ amplitudes are required to fully
specify the state,
irrespective of their particular values: whether some of them have null value 
or such as to make the state maximally entangled. Being absolute, ${\cal M}$ 
depends on the actual ensemble of state vectors, 
and not only the density operator. By definition, a uniform mixture of 2 4-state
systems in the
separable states ${\cal E}_1 \equiv \{|00\rangle, |11\rangle, |22\rangle,
|33\rangle\}$  has the same amount
of entanglement as a uniform mixture of 
entangled states: ${\cal E}_2 \equiv \{
	{(1/\sqrt{2})(|00\rangle + |11\rangle)},
        {(1/\sqrt{2})(|00\rangle - |11\rangle)},
	{(1/\sqrt{2})(|22\rangle + |33\rangle)},
        {(1/\sqrt{2})(|22\rangle - |33\rangle)}
        \}$, namely zero. 
In fact, the two mixtures are indistinguishable, having the same density matrix. 
Yet, we have ${\cal M}({\cal E}_1) = 2\times4\mu=8\mu$ bits and
${\cal M}({\cal E}_2) = 4^2\mu=16\mu$ bits. This reflects the idea that to
nature, viewed as the underlying information processor, all states are pure, and
the entanglement depends on the individual states in an ensemble and not
on the density matrix.

The information processing resource Eq. (\ref{qnmem}) presupposes is very large. Consider 
a ``laptop" of mass $m = 1$ kg and registering
information in the spin of the particles composing it. The total
number of baryons, and therefore the total number $n$ of bits registrable therein, is 
about $n \equiv 1/m_p \approx 6. \times 10^{26}$ bits. The memory ${\cal M}$ that can
support the evolution of this many qubits 
is the superastronomical number $2^{6. \times 10^{26}}\mu \approx
10^{1.81 \times 10^{26}}\mu$ bits, where $\mu > 2$, according to the
computational completeness principle. (It is worth noting that the number
is $10^{(1.81 \times 10^{26})}$, and not the much smaller
$\left(10^{1.81 \times 10}\right)^{26} \approx 10^{260}$; i.e.,
$\log_{10}({\cal M}) \sim O(10^{26})$). 
Not surprisingly, conventional memory in the entire universe, let alone the part 
available to conventional computers, will hardly suffice
to support the information processing overhead corresponding to the evolution of
such a quantum system. Suppose Moore's law has been taken to its possible
extreme, whereby all particles in the universe register a bit each. Estimating the
number of baryons in the universe at $\beta = 10^{76}$ (cf. below), we find 
using Eq. (\ref{qnmem}) that
the most number of qubits ($D=2$) whose state information can be registered is
$n = 76\cdot\log_2(10) - \log_2(64) \approx 246$ qubits. Thus, the entire
accessible world is not enough to track the state of even 300 qubits at just
double precision. In fact, the Bekenstein bound \cite{bek}, together with the
holographic principle \cite{hp}, implies that the maximum information that can be
registered in a system is $A/l_P^2$, where $A$ is the system's area. Applying this
to the universe as a whole implies that the maximum number of (classical) bits that could be
registered by the universe using matter, energy and gravity is $c^2t_U^2/l_P^2
\approx 10^{120}$ bits \cite{llo02}. Solving for $n$ in Eq. (\ref{qnmem}) with ${\cal M}
= 10^{120}$ bits and $\mu=64$ bits
yields 398.6 qubits. In other words, the maximum possible information
registrable in the universe is insufficient to record the state information for more 
than 400 qubits.

Coming to the continuous case, let us consider $n$ 2-state 
particles in a box of volume $V$. 
The memory support required for a single particle
is ${\cal M} = 2(4\pi/3)(ct_U/l_P)^3\mu \approx 10^{182}\mu$ bits, where the wavefunction
has null value for regions outside the box. The memory support for $n$
particles is, as argued above, 
\be
\label{cald}
{\cal M} = \left[2(4\pi/3)(ct_U/l_P)^3\right]^n\mu \approx 10^{182n}\mu~ {\rm bits}.
\ee 
% (Strictly speaking, this is true of strongly interacting particles. In practice, it may
% turn out that some internal degrees of freedom are effectively separable or are of sufficiently
% low energy as to be irrelevant to compute ${\cal M}$, a point clarified in the last parts
% of Sections \ref{vonn} and \ref{eme}.)
It would appear that Eq. (\ref{cald}) in fact applies to distinguishable particles,
with that for indistinguishable particles given by ${\cal M} =
\{2(4\pi/3)(c^3t_U^3/nV_P)\}^n$, since each particle has effectively 
$(4\pi/3)c^3t^3_U/n$ volume to be in.
However, because ${\cal M}$ is an absolute quantity, it keeps track of individual
particles, whether or not they are distinguishable quantum mechanically. Thus, the
former expression is appropriate.
Let us consider the 1 kg laptop. If made up of H atoms, it is constituted of 
$n = 1/m_p \approx 6. \times 10^{26}$ baryons, or about $4n$  
quarks and electrons, assumed to be elementary paticles. 
Thus \mbox{${\cal D} \approx (10^{182})^{4n} = 10^{1.82 \times 10^{29}}$},
so that ${\cal M} \approx 10^{1.82 \times 10^{29}}\mu$ bits. 

It is instructive to contrast this with maximum information ${\cal M}$ that would be
needed, for an `ultimate laptop' weighing 1 kg and occupying 1 liter volume. 
This corresponds to the maximized entropy of the laptop obtained by
converting all its matter into radiation \cite{llo00}. The conventional equation for
blackbody radiation can then be used to estimate the temperature $T$ that would
be obtained if that matter were converted to radiation at temperature $T$:
$\rho c^2 = (\pi^2/15\hbar^{3}c^3)(k_BT)^4$. Solving for the temperature 
gives $k_BT = (15\hbar^3c^5\rho/\pi^2)^{1/4}$. The number
of photons is estimated as $2\rho c^2/3k_BT =
(2/3\ln(2))(\pi^2V/15\hbar^3c^3)^{1/4}E^{3/4}$, which, for the ultimate laptop
yields the value of $n = 1.6 \times 10^{31}$ particles \cite{llo00}. 
We then have
${\cal D} = (10^{182})^n \approx 10^{4.4\times10^{33}}$, so that ${\cal M} =
10^{4.4\times10^{33}}\mu$ bits of memory correspond to the 
quantum state of the ultimate laptop.

Let us consider the action of the 
Hamiltonian $\hat{H} = \sum_{j=1}^n \hbar\omega_j\hat{\sigma}_x^{\otimes j}$
on a system of $n$ qubits initially prepared in the state $|0\rangle^{\otimes n}$,
i.e., each qubit in an eigenstate of the Pauli operator $\hat{\sigma}_z$. Note that,
being separable, it requires in general just $2n\mu$ bits to be specified.
But, after time 
$t$ they will have evolved, by the action of $e^{-i\hat{H}t}$, to the entangled state:
\be
\label{entangl}
|\Psi(t)\rangle = \frac{1}{2^{n/2}}\sum_{k=0}^{2^n-1}\sum_{j=1}^n
e^{-i\omega_j(1-2\pi_k(j))t}|k\rangle,
\ee
where $\pi_k(j)$ is the parity of the first $j$ digits in the binary representation of string
$k$. Thus, we now need $2^n$ (though not all 
independent) amplitudes to describe the entangled
state. It is not difficult to see that, for sufficiently intricate interactions,
all $2^n$ amplitudes will be independent. More importantly, as noted earlier,
the fact of entanglement between the $n$ objects, by itself, irrespective of the 
magnitude of entanglement, is the relevant quantity to quantify ${\cal M}$,
which in this case is $2^n\mu$ bits.

Now suppose that at $t=t_0$, the interaction Hamiltonian is switched off, and the
system $|\Psi(t_0)\rangle$ is subjected to (free) evolution with the Hamiltonian
$\hat{H}_1 = \sum_{j=1}^n \hat{H}_0^{(j)}$. Applying the action
$e^{-i\hat{H}_1t}$ for time $\Delta t$, we find that the state $|\Psi(t_0)\rangle$ evolves 
to:
\be
|\Psi(t_0+\Delta t)\rangle = \sum_{k=0}^{2^n-1}e^{-i(\Omega(k)\Delta t + p(k)t_0)}|k\rangle,
\ee
where $\Omega(k) = n_k(0)E_0 + n_k(1)E_1$, where $n_k(i)$ is the number of $i$'s in
string $k$, and from Eq. (\ref{entangl}), we have $p(k) = \sum_{j=1}^n
e^{-i\omega_j(1-2\pi_k(j))t_0}$. The average energy of the 
state is $\langle \Psi(t_0+\Delta t)|H_1|\Psi(t_0+\Delta t)\rangle = n(E_1 + E_2)/2
= nE$. Each amplitude is evolving with average $\overline{\Omega} =
nE$. This shows that, for the generic entangled state,
the speed of evolution of the whole system is governed by the average energy of the
whole system. Therefore, the computational speed $f_{qn}$ at which a system of $n$
$D$-dimensional objects of average energy $E$ is got by inserting the
appropriate values in Eq. (\ref{qfreq}):
\be
\label{qnfreq}
f_{qn} = \frac{2^{\mu/2}D^nnE}{\hbar} \equiv
\frac{2^{\mu/2}{\cal DE}}{\hbar} ~~{\rm op/s},
\ee
where ${\cal E} = nE$ is the
average energy of the whole system. We can view $f_{qn}$ as the bit rate at which $D^n$
amplitudes of a system with energy $nE$ are being evolved at $\mu$-bit precision.
The clock rate $\omega$ for a system is defined as the computational speed per bit,
given by $\omega = f_{qn}/{\cal M} = 2^{\mu/2}nE/\mu\hbar$. 

The computational speed required to sustain the evolution of the 1 kg laptop of 1
liter is, according to Eq. (\ref{qnfreq}), $f_{qn} = (2^{\mu/2}mc^2/\hbar)\cdot
10^{1.82 \times 10^{29}} \approx 
2^{\mu/2}10^{50 + 1.82\times10^{29}}$ op/s.   
Now, it can be shown that for a matter-dominated universe at its critical density,
the total number of logical operations that can have been performed within
the particle
horizon at time $t_U$ is $(t_U/t_P)^3 \approx 10^{120}$ op/s \cite{llo02}, where $t_P \equiv 
\sqrt{\hbar G/c^5} = 5.391 \times 10^{-44}$ secs. Thus, we find that the entirety of
in-principle accessible operations performed in the universe is insufficient to
support even a second of evolution of the laptop. Put another way, if all the
accessible computation performed in the universe were made available, it would have
sufficed to support the evolution the laptop for no more than $\approx 
10^{120}/(2^{\mu/2}10^{50 + 1.82\times10^{29}}) \approx 2^{-(\mu/2)}
\times 10^{70-1.82\times10^{28}}$ sec, which is much smaller than even 
$t_P$. 

We can similarly compute the bounds for the universe as a whole: 
all matter and radiation in the universe was fashioned from its primordial
content, coming from Milky Way matter, which ultimately originated in the Big Bang
over $10^{17}$ sec ago. Hence, the worldlines of all objects in the universe 
were enmeshed in that distant past, and in principle, the entire universe is a closed,
entangled system. If we subscribe to the standard view that evolution is purely unitary,
then, inspite of the apparent classical behavior of the visible world, there is a
minute amplitude entangling all nearby objects to matter in the distant stars
and beyond. As before, suppose all matter is converted into radiation. The
resulting number $n$ of photons is about $2\rho c^2/3k_BT = 
(2/3\ln(2))(\pi^2V/15\hbar^3c^3)^{1/4}E^{3/4} \approx 10^{90}$, where $\rho =
10^{-27}$ kg/m$^{3}$ corresponding to one proton/m$^3$.
We then have ${\cal D} \approx (10^{182})^n 
\approx 10^{1.82\times10^{92}}$, so that ${\cal M} =
10^{1.82\times10^{92}}\mu$ bits of memory are required to record the state of the
entire universe.
The evolution of the universe is supported 
by a computational speed $f_{qn} = 2^{\mu/2}\cdot 10^{1.82\times10^{92}} E/\pi\hbar =
2^{\mu/2}\times 10^{100 + 1.82\times10^{92}}$ op/s according to Eq. (\ref{qfreq}).

\section{What is nature computing?}\label{compu}

The above sections calculated how many elementary operations and how much memory are
equivalent to the state and evolution of some simple physical systems. As
noted earlier, there are three distinct interpretations of the numbers calculated.
The first interpretation simply states that the number of op/s and number of bits
given here are upper bounds to the amount of quantum computation that are performed
during the course of evolution of a physical system.  
This interpretation should be
uncontroversial: a particle that is passed through a double-slit, for example, can
be regarded as a quantum computer that computes the Fourier transform of the slit
configuration. Since the effect arises because of the interference of
amplitudes, this clearly establishes that, even if the path taken by a
quantum object is not an element of physical reality in the
Einstein-Podolsky-Rosen sense \cite{epr}, computations are performed during 
the evolution of the amplitudes to effect the Fourier transformation.
A more complicated example, involving entanglement, as against single particle
superposition, is the Shor algorithm,
involving the computation of the periodicities of certain modular functions
\cite{shor}. Here again we are reminded in a dramatic way that computations are
performed in all $D^n$ dimensions, even though in a given run, we can access only a
much smaller quantity.

The second interpretation of ${\cal M}$ and $f_{qn}$ notes that the numbers
calculated give a lower bound on the number of bits needed and op/s that must be performed
by a quantum computer that performs a direct simulation of a physical system at
some fixed resolution of ${\cal H}$. 
This interpretation should also be
uncontroversial: it is well recognized that quantum computers are better than classical
computers in certain tasks, specifically simulating quantum systems, because the former 
can compute with massive parallelism determined by the dimension and entanglement 
of the system \cite{fey80}. 
The true of quantification of ${\cal M}$ and $f_{qn}$ will depend
of what the actual degrees of freedom in nature are, which should determine 
the ultimate model of Theory of Everything (TOE) we use, whether it is of
elementary strings \cite{free}, membranes or loops in quantum foam \cite{loop}.

The third interpretation-- that the numbers of bits and op/s calculated here
represent the actual memory capacity and elementary operations performed by
nature-- is more controversial. It is of relevance to novel, computational models of the 
universe such as presented in
Refs. \cite{cah96,zei01,wol02}. 
If this interpretation is meaningful, then an elementary
particle is best viewed simply as an information theoretic entity,
in particular, a {\em data structure}--
contingent regions in space where memory about a single object is
concentrated. We view state information 
as being encoded simply in terms of specifying the wavefunction at each 
point in configuration space, since this is sufficient to reconstruct the entire state of
the particle. The particle's (smeared) position in
configuration space may be regarded as the {\em memory address} for accessing the
information content of the
particle. Interactions emerge in this picture as gates for accessing and/or modifying 
this information content. Interestingly, this picture provides a converse scenario to 
the fact that
virtually all physical interactions can operate as logic gates \cite{llo00}. 

That nature registers an amount of information
equal to the logarithm of the number of all its possible microstates in Hilbert
space seems reasonable, considering that any of them
is a possible outcome of a suitably chosen observable. But
whether or not it makes sense to identify the local evolution of
information-carrying degrees of freedom with elementary quantum logical
operations by nature is a question which the remaining part of this article is
dedicated to. Indeed, we can first ask the question: 
{\em What sort of evidence would be needed to demonstrate an essential 
computational character of nature?} How does one distinguish it from other
{\em effective} models of quantum mechanics? 
We need to clarify what a suitable answer to these questions 
would be. It seems that, to be free from
interpretational and philosophical ambiguity, it must be able to reasonably
demonstrate that certain physical phenomena are better described computationally
than dynamically, eg., the appearance of certain discontinuities could be
attributed to
finite information processing resources supporting the evolution of physical systems.
In the next and following Sections, we argue that quantum measurement provides a 
possible window
onto such a demonstrable essential computational apparatus underlying physical reality.
In particular, as explained below, the classicality of measurement outcomes and of
macroscopic phenomena are manifestations of 
information overflow in this underlying apparatus.
We believe that the computational model of quantum measurement presented here,
in response to the above questions, 
can perhaps induce further insight
into the relationship between physics and computation. 

In such an interpretation: what is nature computing? 
If one chooses to regard nature as performing a computation, most of the elementary
operations in that computation consist of position amplitudes of various
particles constantly evolving, as 
quarks, electrons and photons move from place to place and interact with each
other according to the basic laws of physics. In other words, to the extent
nature is performing a computation, it is ``computing" its own dynamical
evolution \cite{pag82}. Only a extremely small fraction of the universe is classically
accessible, and an even smaller fraction being performed on conventional digital
computers.

\section{Information transitions in large physical systems}\label{mez}

To calculate the number of bits necessary and op/s that are performed during
a measurement requires a model of measurement. In the initial, pre-measurement
step, the measured system $S$ and the measuring apparatus $M$ interact 
via Hamiltonian $\hat{H}$ and become entangled, 
as noted in Eq. (\ref{envent}). For macroscopic systems, this implies
large magnitudes of computational resources to support the required ${\cal M}$ 
and $f_{qn}$. In view of this,
it is not unreasonable in a computational model of physical evolution 
to suppose that there are finite 
parameters and procedures to ensure {\em efficient
and error-free} information processing. The question 
is whether/how such computational parameters and procedures
will be reflected in the overlying dynamics.
As seen earlier, over $10^{1.82 + 10^{92}}\mu$ bits are
required to support the quantum evolution of the universe as an entangled whole. 
In an actual program that stores and accesses so many bits, computational support of
the same order will be needed in the abstract operating system,
introducing more complexity. For example, further working cost is
incurred in creating similar-sized arrays of pointer variables in order to accurately match
the exponentially increasing number of amplitudes to the corresponding multiparty eigenstates.
This demands accurate tracking and error correction. Note that even if nature's
information processing is noise-free in the classical sense (where noise can be introduced
from the external environment), it would seem that 
models of physical reality built on intrinsic randomness, eg.,
Refs. \cite{cah96,zei01}, are expected to require error correction in order to 
counter errors arising purely from random fluctuations, which can introduce Gaussian noise.
Such intrinsic quantum error correction will ensure the observed regularity of the laws of 
physics in nature (in terms of the constancy of the fundamental constants and
laws).

In the following two subsections, we examine two, independent
possible scenarios for the emergence of classicality at a macroscropic level as a
consequence of limitations imposed by the finiteness of the fine-graining of Hilbert
space, which turns out to be equivalent to an effective finite size of the memory and
information processing capacity of the `operating system' underlying physical
systems. The detailed implications of these scenarios
for the dynamics of macroscopic systems is dealt
with in Sections \ref{vonn} and \ref{eme}.

\subsection{Quantum chaos scenario}\label{kaos}
From communication engineering and the study of
natural genetic systems \cite{patl02}, we know that error correction is necessary 
for detection and elimination of errors during information processing. 
It permits the safe processing of input information, and the
transmission of information across space and time. Similarly, 
if the Schr\"odinger evolution of physical systems is indeed a manifestation of
underlying information processing in nature based on intrinsic randomness, 
nature's error correction should
play a role in guaranteeing the temporal and spatial symmetries of physical laws.
For example, in the process physics model
\cite{cah96},  self-referential noise gives rise to the basic structure of space
and some quantum phenomena. Statistically speaking, over $n$ noisy processes, `errors' in 
nature can arise with a probability $n^{-1/2}$. While this will be extremely small,
it can in principle give rise to rare unexpected deviations from the physical laws.
It is very highly unlikely that such errors in
the enforcement of physical laws won't be smudged out;
and, even if detected, they will be counted as a statistical quirk. Yet, in principle,
the role for some sort of error correction can never be ruled out. 
% \cite{arul}
In particular, it will be necessary to thwart the onset of quantum chaos, 
induced by such imperfections, 
even when operating in a closed system,

Now, this may seem quite unexpected considering 
that quantum mechanics is usually known to suppress classical chaos.
Yet, based on the studies of the simulation of a generic model of a quantum computer, 
with qubits interacting with coupling strength $J$,
Georgeot and Shepelyansky \cite{geo02} find that for coupling strength $J > J_c$, 
quantum chaos emerges, induced by imperfections and residual inter-qubit coupling.
It leads to ergodicity of the quantum computer eigenstates.
Here the critical coupling strength $J_c$
is given by $J_c \sim \Delta_0/n$, $\Delta_0$ the average energy distance
between the two levels of a qubit, and $n$ the number of interacting qubits. 
Above the quantum chaos border, an initial register state
$|\psi\rangle$ will spread quickly with time over an exponential number of
eigenstates of the system with residual interaction, destroying the operability of
gates. With the disappearance of non-interacting qubit structure, 
the quantum computer as a whole suffers a melt-down \cite{geo02}. 
In the context of the quantum computer model of the universe, 
this would mean that, because of strong
interactions between particles, an initial random error could be chaotically
amplified until physical reality
sinks into an almost chaotic haze of matter and radiation, in contrast to
the structured world we see around us. Such a chaotic thermalization of
the universe is not conducive to the formation of stable structures, let alone
complex systems like the human beings, as we know them.

In simulations of a quantum computer, the
time scale in the chaotic regime after which quantum chaos sets in is
given by $\tau_X \sim 1/\gamma$, where, as in many-body systems, % \cite{many}, 
the spreading rate $\gamma$ can be estimated as $\gamma \sim nJ^2/\delta$ for $J_c < J
< \delta$ \cite{geo02}. Setting $\delta \approx J$, we write:
\be
\label{tx}
\tau_{\chi} \approx 1/nJ,
\ee
which shows that the number of interacting bodies and the strength of their
interaction induce a more rapid onset of chaos.
In principle, the developement of quantum chaos can be suppressed by
error-correcting codes applied within this time. 
If the error correction code in nature has a
code rate $r$ \cite{coderate}, then ${\cal D} \equiv
D^n$ possible states are encoded into and
decoded from ${\cal D}^{\prime} \equiv D^{n/r}$ vectors. 
From Eq. (\ref{tx}),
the angular frequency at which error correction must be performed is $\omega_{\chi} \ge
nJ/\hbar$. 
The information processing rate for error correction is proportional to
the number ${\cal D}^{\prime}$ of abstract vectors.
However the memory supporting the system, and into which the corrected 
information is ``loaded", is ${\cal M} = \mu{\cal D}$ in number. Therefore, to
stave off potential chaos in physical reality, each supporting bit in the underlying
processor must perform operations at rate $\omega_{\chi} (\mu{\cal D}^{\prime}/{\cal M}) = 
\omega_{\chi}D^{nx}$ where $x = ((1/r) - 1)$ \cite{codrat}. 
It is reasonable to expect that
for error correction to be fast enough to thwart the onset of quantum chaos, the
above rate must not be much larger than the clock rate $\omega$ for the system.
Therefore, the condition that an object's `processor' can correct for
chaotically induced errors fast enough is, $\omega > \omega_{\chi} 
(\mu{\cal D}^{\prime}/{\cal M})$, or:
\be
\label{Xaos}
2^{\mu/2}nE \ge \hbar\omega_{\chi} \mu D^{n}, 
\ee
where we approximate $x \approx 1$.
For fixed $\mu$, the left-hand side increases linearly with $n$, whilst the 
right-hand side increases exponentially.
Clearly, for sufficiently large $n$, the required error correction exeeds
the clock rate available to the physical object. 
Such a large object can no longer process its error
correction quickly enough to stave off the onset of chaos, without 
experiencing a \mbox{`burn-out'}. An object for which the inequality (\ref{Xaos})
is violated is said to be {\em computationally unstable}.

Using Eq. (\ref{qnmem}), we obtain an inequality equivalent to Eq. (\ref{Xaos}): 
\be
\label{gammam}
{\cal M} < 2^{\mu/2}\left[\frac{E}{J}\right] \equiv {\cal M}_1 ~{\rm bits},
\ee
where $E/J$ is the ratio of the average energy to coupling strength for
each object in the system. Effectively, ${\cal M}_1$ can be regarded as the
{\em finite upper limit to the amount of memory available to a system}. 
A sufficiently large, complex system 
for which ${\cal M} > {\cal M}_1$ is computationally unstable because it is trying
to access memory not allocated to it. The situation will lead to a situation
akin to a 
`segmentation fault' \cite{sf} in conventional computing,
which occurs when a computer
program tries to access memory locations not allocated for the program's use, eg.,
when one tries to access an array index which is out of the declared range.

Let us consider an elementary particle, initially in a separable state. 
In this case, $n=1$ and it is assumed that $\mu$ is sufficiently large to
satisfy Eq. (\ref{Xaos}). The particle manifests pure quantum mechanical
behavior, its unitary evolution being governed by Schr\"odinger equation (or an
appropriate relativistic generalization thereof). As the particle
interacts with its immediate surrounding, it rapidly becomes entangled with them,
and $n$ increases. ${\cal M}$ grows exponentially according to Eq. (\ref{qnmem}).
For sufficiently large $n$, inequality (\ref{Xaos}) is violated. 
The system's coherent unitary Schr\"odinger
evolution can no longer be computationally supported for such a large-$n$ system.
Error-correction can no longer be performed
fast enough to ward off emergence of quantum chaos. 
The thus-far computationally stable system attains to computational instability. 

The system's subsequent behavior, which is a key component of the model, is better 
understood computer theoretically than dynamically: the computationally unstable
system experiences the quantum information 
equivalent of a `program crash' \cite{crash}, that
instantaneously projects the system into a separable state of $n$ 
particles. Thereby it re-enforces inequality (\ref{Xaos}) and restores
computational stability, since now $n=1$ for
each of the $n$ disentangled particles.
The memory limit ${\cal M}_1$ can thus be thought of as an automatic
servo-mechanism to stave off the thermalization of the system through the onset of
quantum chaos. The crash of the unitary phase leads to the quantum 
informational equivalent of a  `core dump' \cite{cd}. In our case, 
the {\em core} is generated as a snapshot of one of the $D^n$ separable
states residing in the memory just before the crash, each 
specified with ${\cal M} \sim O(n \times \mu D)$ bits,
corresponding to a {\em mnemonically minimal state}, i.e., a state that, being
separable, requires minimum memory ${\cal M}$ to specify. The core describes the
state of the system after the crash. Therefore, the crash 
represents an abrupt {\em information transition} from a computationally unstable
state to a mnemonically minimal, computationally stable state. The term
`transition' denotes the fact the change is discontinuous, in contrast the
smooth build-up of ${\cal M}$ is the preceding phase of unitary evolution.
And the `core dump' is equivalent to the projection in Eq. (\ref{projex}).
It specifies the initial conditions for the subsequent unitary phase.
Information transition causes the system to factor out momentarily from the rest of 
the universe. Loosely, the system is thereby said 
to have `collapsed' non-unitarily into a random separable state. 

As we find later, the basis set
of mnemonically minimal states into which the unitary phase 
information-transitions, called the transition basis, is under
many circumstances uniquely determined by environment-induced decoherence.
Which particular state in the transition 
basis is chosen is determined probablistically to 
satisfy the Born rule. 
Whether the choice is truely random, or pseudo-random, being determined
by nature's unknown underlying computational architecture, is beyond the scope
of the present work. 
Non-selectively, the information transition turns the
computationally unstable state into a {\em proper} mnemonically minimal mixture.
Thus, information transitions serve as an engine for turning highly entangled
states into proper mixtures.
Upon the crash of the unitary phase of a closed system, memory is
emptied (freed), being made available for the subsequent unitary phase.
We will show in detail in Sections
\ref{vonn} and \ref{eme} that the classical behavior of large systems arises
from the very rapidly alternating sequence of unitary and (information) 
transition episodes. 

Consider $n_{\rm max}$ as the value of $n$ that saturates inequality
(\ref{Xaos}). Taking logarithm on both sides of Eq.
(\ref{Xaos}), we see that, for $n_{\rm max}$ and $\mu$ 
sufficiently large and $J\approx E$, $n_{\rm max} \sim \mu$, i.e., the
number of particles that can become entangled before a system information
transitions,
ie., its wavefunction `collapses', is of the order of the precision with which
Hilbert space is specified.  Roughly speaking, this means that the
number of particles that can get entangled without collapsing is approximately $\mu$.
Or, {\em the scale at which nature becomes classical
reflects the fine graining of Hilbert space}. 
Macroscopic systems, with $n \gg \mu$, will have
their unitary evolution continually interrupted by information transitions, and
hence behave classically. Microscopic systems, with relatively less complexity and
degrees of freedom, for which $n \ll \mu$, being computationally stable, retain
unitary evolution and hence behave quantum mechanically.
For fixed fine-graining of Hilbert space, all systems larger than $n_{\rm max}$
are classical. Provided the universe is large enough, i.e., $n_{\rm max} <
10^{120}$ (according to the estimate that includes gravitational degrees of
freedom \cite{llo02}), classicality at some sufficiently large scale is inevitable.

\subsection{Quantum computational incompleteness scenario}
The main conclusions of the above Subsection follow also without our invoking
quantum chaos, in the following way.
In specific, suppose the state of a system is given by
$|\Psi\rangle = \sum_{k=1}^{\cal D}A_k|k\rangle$, where 
${\cal D}~ (\gg 1)$ is the system's dimension, and
$\sum_{k=1}^{\cal D}|A_k|^2 = 1$. In a uniform superposition, 
$|A_k|^2 = {\cal D}^{-1}$. 
According to the computational completeness principle, 
fine-graining parameter $\mu$ should be large enough to support a uniform
superposition. The maximum
precision at which $|A_k|$, and hence $|A_k|^2$, can be specified is
$2^{-\mu/2}$. Thus, we require $2^{-\mu/2} \leq {\cal D}^{-1}$, or
$\mu \geq 2\log{\cal D}$.
In particlar, if the system consists of $n$ $D$-dimensional objects, then
${\cal D} = D^n$ so that
\be
\label{yfirflo}
\mu \geq 2n\log D,
\ee
for ``high fidelity" evolution. Let $n_{\rm max} = \mu/(2\log D)$ 
be $n$ that saturates inequality (\ref{yfirflo}).
For any fixed $\mu$, a sufficiently large system of entangled objects, 
namely one for which $n \gg n_{\rm max}$, will fail the
computational completeness condition. In such a large system,
an exponentially increasing number ($> (D^n - n_{\rm max})$) of
small ($|A_k|^2 < 2^{-\mu/2}$) but finite amplitudes are being set to zero,
because Hilbert space fine-graining is not fine enough. As a result,
the system can no longer record 
amplitudes and track their evolution accurately enough to enact a 
high fidelity evolution. An object for which the inequality (\ref{yfirflo})
is violated is said to be computationally unstable. This gives the second scenario
for computational instability, apart from that discussed in subsection \ref{kaos}.

Using Eq. (\ref{qnmem}), we can express condition (\ref{yfirflo}) 
for computational completeness as a limit on memory
available to a physical system: 
\be
\label{gammao}
{\cal M} \le \mu\times2^{\mu/2} \equiv {\cal M}_2 ~{\rm bits}.
\ee
This may be compared with the
expression for memory threshold ${\cal M}_1$, obtained in 
Eq. (\ref{gammam}). Since the exponential dependence on $\mu$ will most dominate
both expressions, these two expressions are not inconsistent with each other. In fact,
${\cal M}_1/{\cal M}_2 = E/\mu J$. There is good reason to believe on this
account
that ${\cal M}_1 \ll {\cal M}_2$. The former threshold, being smaller, will be
reached faster, and thereby be the more decisive. However, the ${\cal M}_2$
scenario is simpler, and perhaps more plausible. For this reason, in future
discussions, we will generally refer to ${\cal M}_2$ rather than ${\cal M}_1$.
With this interpretation, we see that the evolution of a computationally
unstable system is analogous to the execution of a conventional program with
invalid pointer de-referencing, which will inevitably lead to a segmentation fault.
A sufficiently large, complex system,
for which ${\cal M} > {\cal M}_2$ is computationally unstable because it is trying
to access memory not allocated to it. 

Let us consider an elementary particle
initially in a separable state, described as a system with $n=1$.
As it interacts with neighboring particles, and become
entangled, it is described as part of a system with increasing $n$.
The memory requirement of this system 
grows exponentially, as given in Eq. (\ref{qnmem}). 
At the point where $n \gg n_{\rm max}$, or
equivalently ${\cal M} \gg {\cal M}_2$, the system becomes computationally
incomplete. Error correction is no longer possible, because the processor 
lacks the Hilbert space fine-graining needed to keep track of the exponentially
diminishing amplitudes.
The effective non-availability of memory for tracking all amplitudes, necessary
for accurate enactment of the multi-particle system's evolution, 
leads to computational instablity.

As in the above burn-out scenario,
the system's subsequent behavior, which is a key component of the model, is better 
understood computer theoretically than dynamically: the computationally unstable
system experiences an information transition, that
instantaneously projects the system into a separable state of $n$ 
particles. Thereby it re-enforces inequality (\ref{yfirflo}) and restores
computational completeness and stability, since now $n=1$ for
each of the disentangled particles.
The memory limit ${\cal M}_2$ can thus be thought of as an automatic
servo-mechanism to safeguard computational completeness.
Indeed, it may not be inappropriate to liken ${\cal M}_2$
to the critical temperature $T_c$ in phase transitioning systems. 
The crash of the unitary phase leads to the quantum 
informational equivalent of a core dump.  
The core is generated as a snapshot of one of the $D^n$ separable
states residing in the memory just before the crash, 
corresponding to a mnemonically minimal state, whose basis is usually determined
by environment-induced decoherence. The crash 
causes the system to factor out momentarily from the rest of the universe. 
The choice of a state in the basis is assumed to be probabilistic, and conforms 
to the Born rule. The chosen state is `uploaded' from the core onto the
physical system, while memory of all other states is irretrievably erased,
as in the crash of a conventional computer program.

As briefly explained in Subsection \ref{kaos}, and to be detailed in Sections
\ref{vonn} and \ref{eme}, macroscopic systems, being computationally unstable,
manifest classical behavior, which arises
from the very rapid alternating sequence of unitary evolution and information
transition. Inasmuch as $\mu = 2n_{\rm max}\log D$, we infer that the
number of particles that can get entangled without suffering information
transition (`collapsing') is approximately $\mu$ (apart from a constant factor).
Thus, we are again led to the physical interpretation that 
{\em $\mu$, the degree of fine-graining of Hilbert space, determines quantum/classical
border.} 

In either scenario presented above,
the ``Heisenberg cut" \cite{joo00} that separates the
microscopic, quantum mechanical realm from the macroscopic, classical one, is
rooted in the information structure of Hilbert space. This is the
new insight that emerges in the computational model. On the conceptual side,
it furnishes a reasonably good suggestion that physical reality is grounded on
informational framework, or, informally, ``an information structure 
lies deeper than physics". Interestingly, this can be interpretted as
conveying the converse of Landauer's aphorism: ``information is physical"
\cite{lan88}!

How large is $\mu$, or equivalently, ${\cal M}_2$? It must be 
large enough to permit the well-tested, purely unitary evolution of small systems that
are effectively isolated during the course of observations, such as photons in
quantum optical experiments and electrons in atoms, that are to good
approximation well shielded from environmental influence. Yet $\mu$ is to be small enough
that macroscopic systems are computationally unstable, and hence, classical.
Here decoherence also plays a role in determining the spatial scales that separate
microscopic and macroscopic regimes. An interesting point is that even if ${\cal
M}_2$ is so small that it cannot support the evolution of more than two
entangled elementary 2-state objects, it would still suffice to support the
separable evolution of all the $\approx 10^{76}$ baryons in the universe, inasmuch 
as $10^{182\times2} \gg 10^{76}10^{182}$ bits.

We can estimate $\mu$, by examining instances of the largest 
objects that manifest coherent quantum wavelike properties. The reason
is that superposition of macroscopic objects really means entanglement of
its constituent objects. Thus, manifestation of wavelike property of a
macroscopic body as a whole means that entanglement can be sustained at that
level of macroscopicity, as quantified by $n$. 
Perhaps the largest 
object (without microscopic degrees of freedom frozen in some way, as in
condensates) that have 
demonstrated wavelike behaviour are fullerene molecules $C_{70}$ \cite{c70}.
Similarly, superposition of nucleotides and amino acids have been invoked to
explain genetic features in terms of the Grover search algorithm \cite{pat01}.
Let us suppose that the largest, isolated object that can unitarily evolve, without
suffering information transition, is an molecule assumed for simplicity to 
consist homogeneously of $n$ 
N (Nitrogen) atoms. Each N atom contains $\approx 50$ elementary particles 
in the form of
quarks and electrons. From Eq. (\ref{cald}), we can compute mnemonic entropy 
as ${\cal D}_{\rm max}\mu = (10^{182})^{50n}\mu = 10^{9100n}\mu$. 
If we assume that the largest isolated organic molecule
that can demonstrate coherent wavelike behavior consists of 1000 N atoms,
then we find ${\cal D}_{\rm max}\mu = 10^{9100000}\mu$, so that
$\mu = 2\log({\cal D}_{\rm max}) \approx 6.0 \times 10^7$ bits, and
${\cal M}_2 \approx 6.0 \times 10^{9100007}$ bits \cite{natbar}.  
This may be compared with the most accurate quantum electrodynamic
experiements performed so far, which have about 36-bit (11 decimal
places) accuracy, and atomic clocks, which have 33-bit (10 decimal places) accuracy. 
In fact, the above calculation only considers the degrees of
freedom according to known physics, which could well be smaller than the actual memory 
requirement. 

The above numerical estimate for $\mu$  
can in principle be updated by looking for larger objects that
demonstrate wavelike behavior, either in 
interferometric set-ups \cite{c70,mar02} or in complex many-body
superpositions \cite{kun03}. Here it is worth noting that the computational
instability must be distinguished from 
the presence of multiple wavelength scales \cite{natbar}, 
quantum chaos and decoherence effects.
% In practice, this may not be easy.
All lead to loss of coherence, albeit for very different reasons. Decoherence
affects open systems, but the rest can occur even in closed systems. 
% of multiple internal degrees of freedom of comparable energy can lead to broad-band
% wavelength scales that wash out interference patterns.
Decoherence can be excluded by performance of a test experiment in consderable
isolation. Careful choice of physical systems and preparation of initial
conditions can help thwart quantum chaos \cite{geo02}.

Some points of comparison of the computational model of quantum measurement
with exisiting objective indeterministic 
models thereof can be made. In the dynamical reduction model \cite{ghi}, 
the state vector reduction is said to be triggered by the spontaneous and 
entirely random collapses of individual 
particles in the state (\ref{envent}), which cascades
the entire entangled system into a separable state. Comparably, in the present computational 
model, an information transition is triggered by a segmentation fault in
the computational framework supporting physical objects. 
In the Penrose model \cite{pen96}, the collapse of the wavefunction is attributed to 
instability of superposition due to gravitational self-energy. Comparably, the
present computational model also connects measurement to the a critical phenomenon,
namely computational instability. The essential
new element in the present model is that the classicality of measurement outcomes
and macro-objects is connected with fundamental information processing properties of nature. 

\subsection{Time-scales}\label{timsca}

The evolution of physical objects in the above scenario is characterized by 
two phases, encountered earlier, and two corresponding time-scales: (1)
the {\em unitary phase}, during which Schr\"odinger (or its appropriate
relativistic generalization) equation strictly
applies, and having duration denoted $\tau_u$; (2)
the {\em transition phase}, during which the unitary phase crashes, i.e.,
the system undergoes information transition,
having duration denoted $\tau_t$. In typical large macroscopic systems, the
rapid cycles of unitary evolution and information transitions
give rise to an apparent continuous classical evolution because
$\tau_u + \tau_c \ll 1$ sec.
Neither $\tau_u$ nor $\tau_t$ is the decoherence time-scale, 
$t_{\rm decoh}$, over which long-distance correlations in physical systems practically
vanish through purely unitary interactions of a system with its environment.

$\tau_u$ can be indefinitely long for isolated, sufficiently small objects, for which no
information phase transitions occur, since ${\cal M} \le {\cal M}_2$ is always satisfied.
Here, no modifications to Schr\"odinger dynamics occurs. This is also
effectively true in macroscopic systems over length-scales smaller than
$\lambda_d$, the decoherence length-scale, % the length 
over which correlations persist
inspite of decohering effects due to the environment.
To estimate $\tau_u$ for systems requiring ${\cal M} > {\cal M}_2$, we begin by defining 
the information length-scale, $\lambda_u$, as the characteristic length-scale of
a region whose volume memory support saturates the inequality (\ref{gammao}). Therefore,
${\cal M}$ for matter within a volume $\lambda_u^3$ satisfies
${\cal M} = {\cal M}_2$. Thus, $\lambda_u$ specifies the lower limit for the length scale
above which matter will not be computationally stable. 
Objects with
lengths scale $L$ such that $L \ll \lambda_u$ ($L \gg \lambda_u$) are called 
microscopic (macroscopic), and will show quantum (classical)
behaviour. Therefore, $\lambda_u$ spatially marks the ``Heisenberg cut" \cite{joo00}, the
mesocopic transition from quantum to classical regimes. In a macroscopic body,
points separated by distance larger than $\lambda_u$ will not be correlated,
because coherent effects do not persist over longer length-scales. Since
decoherence also produces a similar effect, we conclude that the correlation
length scale in a system can not be larger than 
${\rm min}\{\lambda_d,\lambda_u\}$.

Let mean molecular weight $W_a$ for the material making up a system be
given by $W_a \approx 2N_a$, where $N_a$ is mean atomic number.
Then there are $N_a$ electronic and $3W_a = 6N_a$ quark, hence a total of
$7N_a$, degrees of freedom available per atom. 
The number of atoms is $n = \rho\lambda_u^3/W_am_p$.  
It follows from Eqs. (\ref{cald}) and (\ref{yfirflo}) 
that $182 \times 7N_a n\log10 < \mu/2$. 
Solving for $\lambda_u$, we obtain:
\be
\label{lamb0}
\lambda_u < \left[\frac{\mu m_p}{1274(\log10)\rho}\right]^{1/3} \approx 10^{-10}
\left[\frac{\mu}{\rho}\right]^{1/3} {\rm m},
\ee
independent on the atomic specie.
Eq. (\ref{lamb0}) shows that the information length scale becomes smaller at larger
densities, in response to the greater compactness of the degrees of freedom
available. 
For a fluid at $\rho = 1$ gm/cc, with $\mu=10^{12}$ bits,
$\lambda_u$ is about $10^{-7}$ m. 

We estimate $\tau_u$ as the time over which energy can be transmitted across a
length $\lambda_u$, since this is the period over which entanglement can be set up and
memory activated. This implies $\tau_u = \lambda_u/c_s$, where $c_s$ is a characteristic
speed of energy transport in the body. According to these assumptions, $\tau_u$ is
independent of size of a macroscopic body.
Velocity $c_s$ can depend on whether the dominant cohesive energy in the
body is in the form of Van der Waals, covalent, ionic or Hydrogen bonds, or if the 
material is a metal. 
This suggests that metals, where energy is transmitted most efficiently, by
transverse waves set up in the electronic band, are the quickest at becoming
classical. In particular, $\tau_u = \lambda_u/c$. 
A large metallic body suffers information transitions at the highest frequency. Since it
is as yet rather difficult to resolve changes to material properties at such small intervals,
sufficiently large bodies behave effectively classically. For a
metallic body with $\rho = 1$ gm/cc, and $\mu=10^{12}$, we found
$\lambda_u \approx 10^{-7}$ m. 
Then $\tau_u \ge \lambda_u/c = 10^{-16}$ sec.
For C atoms, which have $12\times7=84$ spinorial degrees of
freedom, from Eqs. (\ref{cald}) and (\ref{yfirflo}), we obtain
$10^{182\times84n} = 2^{\mu/2}
\approx 10^{1.5\times10^{11}}$. Solving for $n$, we obtain $n \approx 10^{7}$. Therefore,
for $\mu=10^{12}$ bits, the largest object made of C atoms
that can in principle manifest
coherent quantum behavior, as in the sense of the experiment reported in Ref. \cite{c70}, 
is $10^7$ C atoms. 

Coincidentally, the expression Eq. (\ref{lamb0}) can be obtained by estimating
${\cal M}_2$ in terms of conventional thermodynamic entropy.
The temperature that would be obtained if
matter density $\rho$ is converted completely to photons can be obtained from
the equation for blackbody radiation: $\rho c^2 =
(\pi^2/15\hbar^3c^3)(k_BT)^4$. Entropy density is estimated by $S/\lambda_u^3 
= 4\rho c^2/3T$.
And we estimate memory support for a volume
$\lambda_u^3$ as ${\cal M} = \exp(S/k_B)\mu =
\exp\{(4\pi^2/45)(k_BT\lambda_u/\hbar c)^3\}\mu = {\cal M}_2$, where $\exp(S/k_B)$
is the total number of states available to the system. Solving for $\lambda_u$, we
find:
\be
\label{lamb2}  
\lambda_u < \left[\left(\frac{15}{\pi^2}\right)^{1/4}
 \frac{3\ln({\cal M}_2/\mu)}{4} \right]^{1/3}
 \left[\frac{\hbar}{\rho c}\right]^{1/4} \approx 10^{-10}
\left[\frac{\mu}{\rho}\right]^{1/3} {\rm m}.
\ee

One way to view the smallness of $\tau_u$, is to estimate the spreading
of the wavefunction of macroscopic objects during the unitary phase.
Consider electrons at temperature $T$, with
position uncertainty $\Delta x$ estimated to be their thermal wavelength, i.e.,
$\Delta x = h/\sqrt{m_ek_BT}$ m. Their momentum uncertainty is 
given by: $\hbar/\Delta x$, so that the subsequent spread of the wavepacket as function of
time $t$ is \cite{wigner}:
\be
\label{wig}
\Delta x(t) = \Delta x + \frac{ht}{m_e\Delta x},
\ee
where $m$ is the mass of the body.
The increase in position uncertainty during a unitary phase is
$h\tau_u/m_e\Delta x = \tau_u\sqrt{kT/m_e}$. For a body with $m=1$ gm, and 
at room temperature, assuming $\tau_u=10^{-16}$ sec, we find that
system can spread through no more than $10^{-11}$ m. In actual fact, as we later see,
the localizing tendency of environmental
decoherence will ensure that spreading is much smaller than $\hbar/m\Delta x$,
indeed, even negative.

The transition time-scale $\tau_t$ should be arbitrarily small, or, more precisely,
not larger than $t_p$, Planck time, in order to enforce correlations across
spacelike-separated measurements on entangled systems \cite{sca00}. 
For example, in a single particle Young's double slit
experiment, detection of the particle at some point on the screen should
instantaneously re-set amplitudes at all other points on the screen to zero lest
probability conservation be violated. 
If we interpret the particle's state as information with configuration space
as memory address to access that information (probabilistically), then
the double-slit experiment requires a nonlocal action on the memory, and we require
a speed of information that is arbitrarily large \cite{sca00,zbi00}.
% Similarly, simultaneous measurements of the same arbitrary spin
% observable on two halves of a singlet even separated by lightyears should yield
% strictly correlated results. 
In the case of entangled systems, the nonlocality 
(in the sense of violation of Einstein locality \cite{epr})
follows also from the violation of Bell's inequality \cite{bel64},
considering that the computational model endows the wavefunction with a
subtle, informational realism. Computationally speaking, $\tau_t$ is the duration
over which nature frees memory and dumps the core across the spacelike region
over which the (possibly entangled)
wavefunction is spread. As has been recognized, this
presupposes a tremendous `speed of information' \cite{gar02} and a preferred 
inertial frame \cite{sca00}. 
Nevertheless, 
it can be shown that assumptions of completeness and probablistic nature of measurements
suffice to ensure that any model of quantum mechanics
does not violate no-signalling and Lorentz-covariance at
the level of measurement outcomes even if the speed of information is
superluminal \cite{rem02}. 
\section{Measurements}\label{vonn}

The process of measurement, or, more generally, of intervention by an external
system in the evolution of system $S$, can be described as taking place in three
steps: (a) pre-measurement, in the sense of Eq. (\ref{envent}), whereby
measured system $S$ becomes entangled with apparatus $M$; (b) decoherence,
which is the loss of phase information via interaction and entanglement
of system $SM$ with the environment $Q$ over time $t_{\rm decoh}$
\cite{decoz}; (c) a final, least understood,
 amplification step, where one of the decohered states is obtained as
the ``classical" outcome (non-selectively described by Eq. (\ref{projex})). 

In the conventional description, the step (c) involves discarding the
inaccessible degrees of freedom to obtain a diagonal state in the pointer
basis. Why this cannot by itself complete measurement
was discussed at some length in the 
Introduction, namely that it only leads to improper mixtures, which are 
insufficient to explain selective measurements.
The present computational model differs from the above chain of steps
only in respect of step (c). Specifically, the information transitions
produced via computational instability, introduced 
in the preceding Section, are shown to act as an engine continually turning
improper mixtures to proper mixtures. 
Thereby we are able to eliminate the main
interpretational difficulty that has dogged decoherence as a resolution
to the measurement problem, while preserving all its
quantitative successes.

The von Neumann pre-measurement in Eq. (\ref{envent}) considers a detector with
only one degree of freedom. This would
suggest that the required memory support ${\cal M}
\approx O(\mu D)$, or just $2\mu$ $(\ll {\cal M}_2)$ bits for a qubit,
which is insufficient to induce computational instability. 
A complete description of $M$ and $Q$ involves both macroscopic and microscopic
variables. Let us therefore introduce a complete basis for them, namely
$\{|i,\xi\rangle_M\}$ and $\{|i,\xi,\eta\rangle_Q\}$, where 
$i$ labels a macroscopic subspace in $M$, 
$\xi$ labels microscopic states in that subspace, and $\eta$ those in $Q$
that tend to correlate with $i,\xi$ in $M$. 
It is assumed that the microstates are separable, or near separable (localized
entanglements that are well screened from macroscopic degrees of freedom are
permitted) in the constituent particle states.
We note that there need not be the same number of microstates in all the macroscopic 
subspaces, in $M$ and $Q$, corresponding to a given macroscopic state. We have:
$_M\langle i,\xi|i^{\prime},\xi^{\prime}\rangle_M =
\delta_{ii^{\prime}}\delta_{\xi\xi^{\prime}}$ and
$_Q\langle i,\xi,\eta|i^{\prime},\xi^{\prime},\eta^{\prime}\rangle_Q =
\delta_{ii^{\prime}}\delta_{\xi\xi^{\prime}}\delta_{\eta\eta^{\prime}}$.
	
We consider system $S$ in state $\sum_i a_i|i\rangle_S$, where 
$\{|i\rangle_S\}$ are the eigenstates of the observable that apparatus 
$M$ is supposed to measure. As an extension of Eq. (\ref{envent}), 
the pre-measurement step is given by the more general interaction
between the quantum system $S$, the apparatus $M$
in `ready' state $|r,\xi_0\rangle_M$ and environment in some state 
$|r,\xi_0,\eta_0\rangle_Q$ :
\bex
\label{envent1}
\sum_i a_i|i\rangle_S\otimes|r,\xi_0\rangle_M\otimes|r,\xi_0,\eta_0\rangle_Q
 \stackrel{H}{\longrightarrow} 
|\Psi\rangle &=& \sum_{i,\xi,\eta} a_i\epsilon^M_{i,\xi}
\epsilon^Q_{i,\xi,\eta}
|i\rangle_S\otimes|i,\xi\rangle_M\otimes|i,\xi,\eta\rangle_Q \nonumber \\
&\equiv& \sum_{i,\xi,\eta} a_i|i\rangle_S\otimes|i;\xi\rangle_M
\otimes|i;\xi;\eta\rangle_Q,
\eex
where $\epsilon^M_{i,\xi} = \epsilon^M_{i,\xi}(r,\xi_0)$ and 
$\epsilon^Q_{i,\xi,\eta} = \epsilon^Q_{i,\xi,\eta}(r,\xi_0,\eta_0)$
with the terms $r, \xi_0, \eta_0$ within parenthesis usually omitted for convenience.
We have $\sum_{\xi}|\epsilon^M_{i,\xi}(r,\xi_0)|^2 = 1$ and $|i;\xi\rangle_M$ 
are unnormalized $M$ states whose squared norm gives their statistical weight for
given $i$. Similarly, $\sum_{\eta}|\epsilon^Q_{i,\xi,\eta}(r,\xi_0,\eta_0)|^2 = 
1$ and $|i;\xi;\eta\rangle_Q$
are unnormalized environmental states whose squared norm gives their statistical weight
for given $i$ and $\xi$. 
The basic idea here is that $M$ amplifies the state of $S$, and $Q$ that
of $M$, until, after time $t_{\rm decoh}$, the entangled
environmental states are orthogonal \cite{joo00}, irrespective of microstate
labels $\xi_0$ and $\eta_0$.

Tracing out the environmental degrees of freedom
in Eq. (\ref{envent1}), we obtain (omitting system subscripts):
\be
\label{subsystem}
\rho_{SM} =
\sum_{i,\xi}|a_i|^2|i\rangle\langle i|\otimes |i;\xi\rangle\langle i;\xi|,
\ee
the reduced density operator which possesses the form that we normally associate
with non-selective measurements. The recovery
of the (block-)diagonalized improper mixture (\ref{subsystem}) is conventionally
thought to complete
measurement. While this encounters some
interpretational difficulty, nevertheless the fact remains that
the dynamical description of subsystems does not depend
on whether they are given by of proper or improper mixtures. 
Such reduced dynamics has been
studied by various authors, starting with Lindblad \cite{lind} and Gorini,
Kossakowski and Sudarshan \cite{gor76}. 
In particular, let $\rho$ be the reduced density obtained after environmental degrees
of freedom have been traced out. The environment is unknown and assumed to produce
random walks of the state vector that are rapid compared to the evolution
generated by the system's base Hamiltonian $H_0$. Thus $\rho$ undergoes a Brownian
motion superposed on an ideal motion, the result of which can be shown to follow
the Lindblad equation:
\be
\label{eqlind}
\hbar\frac{\partial\rho}{\partial t} = i[H_0,\rho] -
\frac{1}{2}\sum_k\left(\hat{L}_k^{\dag}\hat{L}_k\rho + \rho \hat{L}_k^{\dag}\hat{L}_k - 
2\hat{L}_k^{\dag}\rho \hat{L}_k\right),
\ee
with arbitrary generators $\hat{L}_k$ in ${\cal H}$ \cite{lind}.
These observations, and the main conclusions that follow below, are not affected by
weakening some of the assumptions that go into the model of measurement given
above: for example, the states of $M$ and $Q$
can in general be impure, and $Q$ states need not be coupled to $M$'s microstates.

The Bassi-Ghirardi theorem \cite{bas00}
implies that von Neumann's {\em chain of 
efficient causes} in the course of measurement cannot be terminated 
without departure from linear superposition, unless we admit a many-worlds
interpretation \cite{joo00}. In the computational
model, this chain is terminated at finite depth, when sufficient 
entanglement has accumulated in the $SMQ$ system so that 
${\cal M} = {\cal M}_2$. When the ${\cal M}_2$ bound
is exceeded, the physical system becomes computationally
unstable, and information transitions into a
proper mixture. Thus, the main theme here
is that: {\em Information transitions induced by computational instability
continually drive (mixtures of) improper mixtures into (mixtures of) proper
mixtures.} 
Often, the effect of information transitions can be treated as
a continuous process, even though they are jerky, unlike decoherence.
Let us estimate ${\cal
M}$ induced by a $D$-dimensional obervable $\hat{A}$ of $S$. The pointer on $M$, assumed 
to be made of silicon, 
must have $D$ macroscopically distinguishable positions. Suppose the pointer 
weighs 1 gm, and thus has about $n=3.5 \times 10^{22}$ Si atoms, or
$7n = 2.5 \times 10^{23}$ spinor degrees of freedom. 
According to Eq. (\ref{cald}), an estimate for memory
is ${\cal M}(SM) = D{\cal D}\mu = D\times10^{182\times7n}\mu \approx
10^{3.0\times10^{26}}D\mu$ bits.
Clearly, even though possibly dim$(\hat{A}) = D \approx O(1)$, measurement of
$\hat{A}$ results in a rapid build-up 
of very large ${\cal M}$. 
Let $t_{\rm decoh}$ be the time over which a system decoheres, i.e., over which 
$\rho_{SM}$ becomes spatially block-diagonalized on account of interaction
with the environment. Suppose $\tau_u \ge t_{\rm decoh}$. 
This implies that ${\cal M}$ is sufficiently large, so that the information transitions 
come after environmental decoherence in the observational chain of events. 
(Later we shall find that the main conclusions follow even if $\tau_u < t_{\rm decoh}$).

With these assumptions, consider the system $S$ at $t = 0$ in the initial state
in (\ref{envent1}), when
it is brought in contact with $M$ for measurement. As the atoms on
$S$ interact with those in $M$, and the latter with atoms in $Q$,
the resultant entangling effect propagates into the environment until, at
$t=\tau_u$, because ${\cal M} = {\cal M}_2$, there is a memory overflow in
the combined system $SMQ$. Consequently, it 
becomes computationally unstable and suffers 
an information transition (`a crash'), whereby
of the larger than $10^{(3 \times10^{26})}D$ entangled possibilities, a
single random state $|i\rangle|i,\xi\rangle|i,\xi,\eta\rangle$ is selected, with
probability $|a_i\epsilon^M_{i,\xi}\epsilon^Q_{i,\xi,\eta}|^2$, with information of
all other states being irreversibly erased from the system's memory.
Memory thus freed is
now available for further Schr\"odinger evolution in the succeding unitary phase. 
The probability that outcome $i$ is obtained on the measured system $S$ is
$|a_i|^2 = \sum_{\xi,\eta}|a_i\epsilon^M_{i,\xi}\epsilon^Q_{i,\xi,\eta}|^2$. Thus,
we recover the usual Born rule for projection. Thereby, non-selectively
speaking, information transition results in a {\em proper ensemble of
$S$ states correlated with the pointer basis elements of $M$}, starting from
the improper mixture of $SM$ at the end of the preceding unitary phase. The particular
state resulting from the information transition serves as initial condition for the
next unitary phase that continues briefly until the next transition phase. 
The evolution of macroscopic systems consists of such repeated cycles of unitary and
transition phases. 

Considering that measurement outcomes are non-superpositional in the
pointer-basis, 
the duration $\tau_u$ of the unitary phase cannot
be larger than the time over which the value of the measured observable of
$S$ is read out from $M$, i.e., $\tau_u < t_{\rm read:out}$. Typically,
$t_{\rm read:out} \le 10^{-3}$ sec. Therefore, $\lambda_u$ can in principle be
as large as $10^{-3}c = 10^5$ m, so that according to Eq. (\ref{lamb0}),
$\mu > 10^{45}$ bits. We note that although the $SMQ$ system could have
decohered much earlier, only the final amplified read-out guarantees that the
$SM$ mixture is proper, i.e., the particular read-out state is pure.
It is important to note that,
the mechanism of decoherence assures us that, post-measurement, the
pointer position at some $i$ is robust because, under fluctuations due to
quantum uncertainty or more conventional environmental action
(namely, noise), it migrates 
only over $\xi$ in a subspace $\{|i,\xi\rangle\}$ labelled by fixed $i$
\cite{joo00}. The pointer state is therefore only passively recognized by the
environment. Indeed, a measuring device would be useless if its pointer states were
not stable against decoherence.

Non-selectively, one can view the state $\rho_{SMQ}^{\prime}$ of the system
just after information transition as obtained as if the environment 
locally `measures' the state $\rho_{SMQ} \equiv |\Psi\rangle \langle\Psi|$ 
in the $\{|i\rangle|i,\xi\rangle|i,\xi,\eta\rangle\}$ basis.
Here $|\Psi\rangle$ is given by Eq. (\ref{envent1}).
We call this basis in $Q$, which determines the ensemble into which the system
collapses upon undergoing information transition, as the `transition basis'.
`Measurement' in this leads to the `transition ensemble', 
denoted ${\cal E}_1 \equiv \{|a_i|^2|i\rangle|i;\xi\rangle|i;\xi;\eta\rangle\}$,
whose density operator is given by the block-diagonal matrix:
\be
\label{decohr}
\rho_{SMQ}^{\prime} =
\sum_{i,\xi,\eta}|a_i|^2|i\rangle\langle i|\otimes |i;\xi\rangle\langle i;\xi|
\otimes|i;\xi;\eta\rangle\langle i;\xi;\eta|.
\ee
In this juncture, the state of $\rho_{SMQ}^{\prime}$ momentarily carves 
out into a separable state from the rest of the universe. 
Therefore, non-selectively speaking, the system $SMQ$, and indeed
the universe, {\em even as a whole}, now exist in a mixed state.
In contrast, in an entirely unitary evolution scenario, the final 
state of $SMQ$ remains 
a pure, entangled state of the form in Eq. (\ref{envent1}). 

Inspite of this, it is easy to see that {\em the computational model 
implies  the same reduced dynamics as unitary Schr\"odinger evolution}. 
This follows from the result that the reduced density
operator of a subsystem is unaffected by local operations on another subsystem
\cite{nc,nota}. In particular, the dynamics of the $SM$ subsystem is independent of
`measurements' and the choice of the transition basis in $Q$, i.e.,
\be
\label{crux}
{\rm Tr}_Q\left(|\Psi\rangle\langle\Psi|\right) = 
{\rm Tr}_Q\left(\rho^{\prime}_{SMQ}\right).
\ee
This means the master equation for the subsystem $SM$ is the same, whether or not
the system as a whole is subjected to information transitions. 
In general, this
implies that the evolution of $\rho$, the reduced density matrix for a subsystem 
not isolated from the environment, is governed by the Lindblad
equation (\ref{eqlind}) even in the computational model. The new element lies in
the interpretation: $\rho$ in the conventional description is an
improper mixture of states, whereas 
$\rho$ is a {\em proper} mixture of pure states in the computational model.
As a result, in the latter model
there arises no difficulty in reconciling the observed purity of the 
final outcome on $S$ in a given individual measurement with the final state obtained 
according to the formalism. 
The model thus resolves the measurement problem. 

In the context of the famous Schr\"odinger cat paradox \cite{schr}, we can regard the
hapless cat, or more precisely the state of feline life or death, as pointer
states to some quantum (emission) process. The act of entangling the process
with the cat generates an immense entanglement. The resulting 
computational instablity causes the initial coherent superposition of the
process to become a proper incoherent mixture, classically correlated with the
cat's state.

Measurement in a purely unitary dynamics can in principle be reversed by suitably
accessing environmental degrees of freedom, and in general those of the entire
universe. In the computational model, they are in principle irreversible.
As these degrees of freedom are
fairly inaccessible, and most realistic measured systems are open, the computational 
model is practically indistinguishable from standard quantum theory in this
respect. All the essential above arguments
for the resolution of the measurement problem in the computational model 
hold good even if $\tau_u < t_{\rm decoh}$, since reduced dynamics is independent
of the propriety (status of being proper or improper) of the mixture. However, as
shown below, only at $t > t_{\rm decoh}$ does a preferred transition basis
for the environment, determined by decoherence, emerge.

The discussion above presumes that the transition ensemble of states
$\{|i\rangle|i,\xi\rangle|i,\xi,\eta\rangle\}$ is determined by pre-measurement 
and decoherence, given by (\ref{envent1}). This implies that
the triple basis in the $SMQ$ system, comprising of 
$\{|i\rangle_S\}$ (the eigenstates of the measured observable $\hat{A}$),
$\{|i,\xi\rangle_M\}$ (the pointer (micro-)states),
and $\{|i,\xi,\eta\rangle_Q\}$ (the environmental eigenstates correlated with
$M$ states) is
the preferred basis for information transition, reflecting the well-known
``(preferred) basis problem" \cite{basprob}: that measurement leaves the system as 
definite in a preferred basis. We denote this basis for
${\cal H}_S \otimes {\cal H}_M \otimes {\cal H}_Q$, correlated with
$M$'s pointer states,
as the `extended pointer basis'. The measurement problem can then be posed thus:
why is the extended pointer basis superselected to be the transition basis? 
The answer arises naturally in the computational model.
Since the system's information
transition occurs because of memory overflow, it is reasonable that
the ensemble it leads to should on average require minimal memory support.
In the above example, the reduction in
memory support from state (\ref{envent1}) to some random separable state
$|i\rangle|i,\xi\rangle|i,\xi,\eta\rangle$ is exponential-- to be precise, ${\cal D}$-fold. 
Now, if the system is completely decohered, the states $|i,\xi,\eta\rangle_Q$
will be almost orthogonal to each other, and the transition mixture will contain
only states separable in the tripartite system
$SMQ$, namely elements from the extended pointer basis. 
Let $D \equiv {\rm dim}({\cal H}_S)$, and $M$ be constituted of $\overline{m}$ 
elementary particles, $Q$ of $\overline{q}$ elementary particles. Then, the
memory support required for the right hand side state in Eq. (\ref{envent1}) is
${\cal M}(|\Psi\rangle) = D10^{182(\overline{m}+\overline{q})}\mu$.
In ensemble ${\cal E}_1$, $S$ and $M$ (and its constituents, given the microscopic
correlations) exist in separable states, so that
memory support required for any state in ${\cal E}_1$ is 
${\cal M}({\cal E}_1) = (D+(\overline{m}+\overline{q})10^{182})\mu \ll
{\cal M}(|\Psi\rangle)$.
Here it is worth recollecting that ${\cal M}$ is an absolute quantity. The
exponentially large difference between ${\cal M}({\cal E}_1)$ and 
${\cal M}(|\Psi\rangle)$ holds inspite of Eq. (\ref{crux}).

An information transition into any other basis leads to entanglement
in the transition ensemble, and hence on average
larger memory requirement. Suppose the transition
basis is \mbox{$|i^{\prime},\xi^{\prime},\eta^{\prime}\rangle_Q \equiv
\sum_{i,\xi,\eta}
U^{i\xi\eta}_{i^{\prime}\xi^{\prime}\eta^{\prime}}|i;\xi;\eta\rangle_Q$}, where
$U^{i\xi\eta}_{i^{\prime}\xi^{\prime}\eta^{\prime}}$ is a unitary transformation
in ${\cal H}_Q$, satisfying $\sum_{i,\xi,\eta}
{U^{\dag}}^{i\xi\eta}_{i^{\prime}\xi^{\prime}\eta^{\prime}}
U^{i\xi\eta}_{j^{\prime}\phi^{\prime}\gamma^{\prime}}
= \delta_{i^{\prime}j^{\prime}}\delta_{\xi^{\prime}\phi^{\prime}}
\delta_{\eta^{\prime}\gamma^{\prime}}$.
The eigenstates $\{|i^{\prime},\xi^{\prime},\eta^{\prime}\rangle_Q\}$ 
are not correlated with the pointer basis of $M$. 
At the end of the pre-measurement stage, we can rewrite the final state of 
Eq. (\ref{envent1}) as:
\be
\label{envent2}
\sum_{i^{\prime},\xi^{\prime},\eta^{\prime}} 
\left(\sum_{i,\xi,\eta}
U^{i\xi\eta}_{i^{\prime}\xi^{\prime}\eta^{\prime}}
a_i|i\rangle_S\otimes|i;\xi\rangle_M\right)
\otimes|i^{\prime};\xi^{\prime};\eta^{\prime}\rangle_Q,
\ee
The ensemble (of unnormalized states) that would result from an information transition
wherein $Q$ `measures' in the primed basis is denoted \mbox{${\cal E}_2 \equiv
\left\{\left(\sum_{i,\xi}
\alpha_{i,\xi}(i^{\prime}\xi^{\prime}\eta^{\prime})
a_i|i\rangle_S\otimes|i;\xi\rangle_M\right)
\otimes|i^{\prime};\xi^{\prime};\eta^{\prime}\rangle_Q\right\}$},
where $\alpha_{i,\xi}(i^{\prime}\xi^{\prime}\eta^{\prime}) \equiv 
\sum_{\eta}U^{i\xi\eta}_{i^{\prime}\xi^{\prime}\eta^{\prime}}$.
We note that ${\cal E}_2$ is an ensemble of $SM$ {\em entangled} states classically
correlated with $Q$ states. As the density operators
corresponding to the two ensembles are identical, i.e., $\rho({\cal E}_1) =
\rho({\cal E}_2)$, the two ensembles are locally indistinguishable, and 
quantified by the same amount of von Neumann
information and of quantum entanglement, as quantified using 
any suitable generalization of conventional 
methods applicable to pure states \cite{bruss}.

In contrast, state information ${\cal M}$, being 
an absolute informational measure, depends on the  
{\em ensemble} of states, rather than on the density matrix.
In ensemble ${\cal E}_2$, the system $SM$ exists in an entangled
states with very large Schmidt ranks. On the other hand, $Q$ exists in a
definite microstate, with its constituents in a separable state:
\mbox{${\cal M}({\cal E}_2) = D\times10^{182\overline{m}} \times
\overline{q}10^{182}\mu = D\overline{q}10^{182(\overline{m}+1)}\mu
\gg {\cal M}({\cal E}_1)$}.
Thus, ensemble ${\cal E}_2$ is not mnemonically minimal even though
${\cal M}({\cal E}_2) \ll {\cal M}(\Psi)$ . 
As the information transition aims to free the segmentation faulted
memory supporting a physical system, the transition basis and
transition ensemble that minimize state information will be preferred.
In particular, the {\em mnemonically minimal mixture} ${\cal E}_1$, that 
corresponds to the extended pointer basis in (\ref{envent1}), 
is the preferred outcome of information transition. Any other basis, and the
resultant ensemble ${\cal E}_2$ is
dispreferred as it does not enable the largest freeing of memory.
This can be stated as a fundamental superselection principle 
(`mnemonic minimization principle'): {\em the transition-basis
is given by one that minimizes state information}. Therefore,
decoherence helps select the preferred basis to play
the role of the transition-basis for the $SMQ$ system, in this
case the extended pointer basis. Information transitions play the complementary
role of converting improper mixtures to proper mixtures in this selected basis.
Thus, in conjunction with decoherence, the computational model resolves the basis
problem.

Actually, in the above measurement process,
 the role of the decohering environment is sufficient, but not
necessary, to explain measurement. Merely the assumption 
$\langle i,\xi|i^{\prime},\xi^{\prime}\rangle = \delta_{ii^{\prime}}
\delta_{\xi\xi^{\prime}}$ and the mnemonical minimization principle suffice
to obtain the mixture Eq. (\ref{subsystem}). With these assumptions, the role of
$Q$ is not very different from the role of $M$. However, we include $Q$ because
in fact interaction with the environment is inevitable. Furthermore,
environmental
decoherence plays another important role: it ensures that the extended pointer
basis indeed consists of localized wavepackets \cite{joo85}. 
As a result, the states in the
transition ensemble also correspond to localized wavepackets,
which is essential to explain why measuring apparatuses and their pointers
manifest particle-like localization.
Decoherence in the present model
thus plays the 
twin important roles of (1) helping determine the transition-basis; (2) 
localization of decohered states. 

If $\tau_u < t_{\rm decoh}$, the system has not entirely decohered during
early information transitions,
and the environmental states correlated with the $|i\rangle|i,\xi\rangle$'s in
Eq. (\ref{envent1}) 
are not the mutually orthogonal vectors $\{|i,\xi,\eta\rangle\}$, but a set of
non-orthogonal vectors, which we denote by $\{|\underline{i};\underline{\xi};
\underline{\eta}\rangle\}$.
The mnemonic minimization principle still
implies a unique transition ensemble: ${\cal E}_1^{\prime} 
\equiv \{|a_i||i\rangle|i;\xi\rangle|\underline{i};\underline{\xi};
\underline{\eta}\rangle\}$, with
no preferred environmental basis. 
Ensemble ${\cal E}_1^{\prime}$ may be considered as being obtained when an
incompletely decohered version of state $|\Psi\rangle$ in Eq. (\ref{envent1}) is
`measured' by $M$ in the $\{|i,\xi\rangle\}$ basis. Hence, in place of
$\rho_{SMQ}^{\prime}$ in Eq. (\ref{decohr}), we get:
\be
\label{decohr0}
\rho_{SMQ}^{\prime\prime} =
\sum_{i,\xi,\eta}|a_i||i\rangle\langle i|\otimes |i;\xi\rangle\langle i;\xi|
\otimes|\underline{i};\underline{\xi};\underline{\eta}\rangle\langle \underline{i};
\underline{\xi};\underline{\eta}|.
\ee
This makes little difference to measurement observer
because ${\rm Tr}_Q(\rho_{SMQ}^{\prime\prime}) = 
{\rm Tr}_Q(\rho_{SMQ}^{\prime}) = 
{\rm Tr}_Q\left(|\Psi\rangle\langle\Psi|\right)$.
Once the system fully decoheres, which is quite 
rapid \cite{decoz}, the extended pointer basis, built up from localized
particles, emerges as the preferred basis. In particular, considering that
pointers are localized and the read-out is classical, we expect that 
$t_{\rm read:out}$, the minimum time required for $M$'s measurement result to be
read out, satisfies $t_{\rm read:out} > {\rm max}(\tau_u, t_{\rm
decoh})$. Thus, together with decoherence, the computational
model resolves the measurement problem and the basis problem in a way that agrees
with classical experience and quantum theory as applied to 
well-tested quantum phenomena.

In constrast to almost all other models for quantum measurement
(save perhaps the many-worlds interpretation), 
the present computational model can be said to introduce minimal new physics,
according to two criteria. 
One is that, the computational model is identical with Schr\"odinger
evolution for microscopic systems, and implies the same subsystem
dynamics as Schr\"odinger evolution for open systems. Microscopic systems
include isolated objects like photons in space or in quantum optical set-ups.  
Subatomic particles like electrons in atoms that are well screened from
getting strongly entangled with macroscopic degrees of freedom and whose 
potentials have length-scales much smaller than 
${\rm min}\{\lambda_d,\lambda_u\}$ are also effectively unaffected
by information transitions and behave quantum mechanically. Open systems include most
macroscopic systems, such as measuring apparatuses. Thus, the computational
model is identical/consistent with quantum dynamics in all the relevant and
accessible regimes in which quantum mechanics has been tested.
The second reason is that information transitions,
which essentially distinguish the computational model from standard
Schr\"odinger evolution, are formally like familiar measurements. Moreover, 
information transitions are not dynamical processes, in
the sense following sense.
As pointed out in Section \ref{timsca}, the transition in Eq. (\ref{envent1})
is almost instantaneous
across configuration space. It cannot be described dynamically within the spacetime
arena, except possibly in a pre-quantum theory for the
computational architecture underlying information transitions, which is beyond
the scope of the present work. We believe information transitions are best
understood as such: as computational procedures, that represent a window onto
nature's basic information theoretic framework.

One point worth noting is that it is not just the size 
of the objects involved, but the number of degrees of freedom
becoming available % and entangled 
during interaction, 
that determines whether or not the interaction will lead to
information transitions. This may mean, for example, that certain internal degrees of
freedom may be separable or contain sufficiently low energy, that they may be
irrelevant to compute ${\cal M}$. A simple illustration of this is found in the fact that
it is absorption rather than deflection or rotation (of the polarization of) of photons 
that plays a more important role in engendering information transition in optical systems.
Indeed, a {\em massive} mirror or lens preserves superpositions but
a {\em small} detector does not. The latter is the stronger 'state reductor'. 
It is not simply the mass of the optical element, but also the operation, that 
counts. In reflection or refraction, an electron excited by the incoming 
photon, or the atom to which the electron transmits the photon, re-transmits the 
photon elastically. As a result, the passage of the
photon does not create entanglement among electrons and atoms along its path.
Non-absorptive photon manipulations do not consume much memory, and do not lead to 
information transitions. They preserve
the coherence of the incoming beam.
In contrast, absorption leads to an unpredictable scattering,
excitation and interaction of particles along the photon's path, which
leads to large entanglement and thence information transition.
As a result, even a
tiny photon-detector is a more efficient state reductor than a massive mirror.
A rigorous study of the photon's scattering and its absorption, 
and thence the correct estimate for ${\cal M}$,
should take into consideration the many possible configurations
in which heat can be dissipated through and from a detector. 
Now, scattering is also known to be an efficient decoherer \cite{joo85,joo00}. We thus
obtain the thumb rule that processes that increase entropy and lead to strong
decoherence are also efficient at inducing information transitions.

\section{Emergence of classicality in the macroscopic world}\label{eme}

As in other models of measurement, classicality of the macroscopic 
world is closely related to the fact that outcomes of quantum measurements are classical.
Closed, small systems, for which ${\cal M} < {\cal M}_2$ or
whose size $L$ satisfies $L < \lambda_u$, are entirely quantum mechanical. 
Their dynamics is captured by Schr\"odinger equation, or the appropriate relativistic
generalizations thereof. 

On the other hand, a macroscopic body, for which $L > \lambda_u$, will evolve
naturally into a highly entangled system in terms of its basic degrees of freedom,
leading to computational instability.
The resultant rapid cycles of information transition and unitary evolution
over a timescale of $\tau_u$ 
cause it to behave classically.
Its state is described by a proper mixture or, alternatively, probablistic evolution of
a pure state. We note that, even in macroscopic bodies,
subatomic particles like photons in quantum optical experiments, and
electrons in atoms, that are well screened from
macroscopic degrees of freedom and whose
potential varies significantly over distance 
${\rm min}\{\lambda_d,\lambda_u\}$ are also effectively unaffected
by information transitions and hence behave quantum mechanically. 
Macroscopic systems $Z$ that we usually encounter are open systems,
decohered by environment $Q$, which leads to 
to the localization of its constituent particles. A macroscopic open system 
is thus described as a proper mixture of localized particle states. We will use this
insight to see why Schr\"odinger cat states (macroscopic systems existing in
coherent superpositions of macroscopically distinguishable states) are forbidden.

The correct quantitative way to determine
whether a system is computationally unstable
will depend on the ultimate constitution, and hence true degrees of freedom,
of matter. However, we can hope to get
approximately correct answers if we compute ${\cal M}$ and ${\cal M}_2$ using the
same model for the microscopic constitution of matter. In so far as the chosen
model of material constitution is not basic, both ${\cal M}$ and ${\cal M}_2$ will be
similarly underestimated. 

It is well acknowledged that a macroscopic object like a cat 
does not exist in a Schr\"odinger cat \cite{schr} state like: 
\be
\label{falsk}
|{\rm \upsilon}\rangle = \frac{1}{\sqrt{2}}(|{\rm here}\rangle
+ |{\rm there}\rangle).
\ee
As in the preceding Section, we introduce the various microstates $\xi$ into the
macroscopically distinguishable states of spatial proximality and distality, namely, 
$\{|{\rm here},\xi\rangle, |{\rm there},\xi\rangle\}$, satisfying
$\langle {\rm here},\xi| {\rm here},\xi^{\prime}\rangle =
{\langle {\rm there},\xi| {\rm there},\xi^{\prime}\rangle =
\delta_{\xi\xi^{\prime}}}$, and
$\langle {\rm here},\xi| {\rm there},\xi^{\prime}\rangle = 0$.
Over a period of time $t_{\rm decoh}$, the
environmental states the cat is entangled with become orthogonal
states $\{|{\rm here},\xi,\eta\rangle,
|{\rm there},\xi,\eta\rangle\}$ satisfying
$\langle {\rm here},\xi,\eta| {\rm here},\xi^{\prime},\eta^{\prime}\rangle =
\langle {\rm there},\xi,\eta| {\rm there},\xi^{\prime},\eta^{\prime}\rangle =
\delta_{\xi\xi^{\prime}}\delta_{\eta\eta^{\prime}}$, and
$\langle {\rm here},\xi,\eta| {\rm there},\xi^{\prime},\eta^{\prime}\rangle = 0$.
With these definitions, in place of Eq. (\ref{falsk}) we have
for the state of the combined cat+environment system:
\be
\label{trygg}
|\Upsilon\rangle = \sum_{\xi,\eta}\frac{1}{\sqrt{2}}\left(\alpha_{\xi,\eta}
|{\rm here},\xi\rangle|{\rm here},\xi,\eta\rangle ~+~ 
\beta_{\xi,\eta}
|{\rm there},\xi\rangle|{\rm there},\xi,\eta\rangle\right)
\ee
where 
$\sum_{\xi,\eta}|\alpha_{\xi,\eta}|^2 = \sum_{\xi,\eta}|\beta_{\xi,\eta}|^2 = 1$.
We estimate that a 1.4 kg cat 
contains about $n = 1.4/14m_p \approx 10^{26}$ N atoms (on average), or $7n$
basic, quark and electronic degrees of freedom. This
corresponds to ${\cal M}_{\rm cat} \approx \left(10^{182}\right)^{7n}\mu =
10^{1.2\times10^{29}}\mu$  bits to represent the cat alone. The total
mnemonic support ${\cal M}_{\rm tot}$ for the cat+environment system is given by
$\log{\cal M}_{\rm tot} \approx \log{\cal M}_{\rm cat} + \log{\cal M}_{\rm env}$, 
where ${\cal M}_{\rm env}$ is the memory
needed for the unknown (but immense) environment. It is reasonable to
expect that ${\cal M}_{\rm env} \gg {\cal M}_{\rm cat}$. 

From the observed classicality of cats, it follows that
the state $|\Upsilon\rangle$ in Eq. (\ref{trygg}) requires
${\cal M} > {\cal M}_2$ bits to specify. Such a state,
being computationally unstable, cannot survive for much longer than
$\tau_u$ sec, after which time it suffers an information transition. 
It repeatedly undergoes such transitions that interrupt
unitary phases lasting about $\tau_u$ sec. 
After time $t \ge t_{\rm decoh}$,
decoherence superselects a preferred cat+environment
basis, consisting of localized wavepackets,
given by the basis in which Eq. (\ref{trygg}) is written, which
emerges as the transition basis.
Loosely, one could say that in a given episode of information transition,
the environment `measures' the state $|\Upsilon\rangle$ in the basis
$\{|{\rm here},\xi,\eta\rangle, |{\rm there},\xi,\eta\rangle\}$, 
information-transitioning the cat into a specific
microscopic state of the cat, say a $|{\rm here},\xi\rangle$ with probability
$\sum_{\eta}|\alpha_{\xi,\eta}|^2/2 \equiv |\alpha_{\xi}|^2/2$, or a 
$|{\rm there},\xi\rangle$ with probability
$\sum_{\eta}|\beta_{\xi,\eta}|^2/2 \equiv |\beta_{\xi}|^2/2$. Non-selectively speaking,
the cat is transformed from state (\ref{trygg}) into one of the states in the
transition ensemble ${\cal E}_1 = \{|{\rm here};\xi\rangle, |{\rm
there};\xi^{\prime}\rangle\}$ given by the {\em proper} mixture:
\bex
\label{mixcat}
\rho_{\rm cat} &=& \frac{1}{2}\sum_{\xi}(
|\alpha_{\xi}|^2|{\rm here},\xi\rangle\langle {\rm here},\xi| +
|\beta_{\xi}|^2|{\rm there},\xi\rangle\langle {\rm there},\xi|) \nonumber\\
  &\equiv& \frac{1}{2}\sum_{\xi}(
|{\rm here};\xi\rangle\langle {\rm here};\xi| +
|{\rm there};\xi\rangle\langle {\rm there};\xi|).
\eex
This entails the interpretation that the cat actually exists in one of the pure states 
that enter into the mixture (\ref{mixcat}). Decoherence ensures that the macroscopically
distinguishable states of $|{\rm here}\rangle$ or $|{\rm there}\rangle$ are robust
and made up of localized particles. 
After a given information transition, the cat 
evolves only into superpositions of microstates corresponding to one of these
macrostates, before being reduced into one of those microstates by the succeeding
information transition, and so on. 
In this way, the familiar classical cat emerges
as the combined effect of decoherence and information transitions.

The effect of large numbers of interacting degrees of freedom on the decay of macroscopic
superposition \cite{mya00} is central to the present model as it is to most
models of measurement.
We can summarize this line of argument for the classicality of the macroscopic world 
as follows: macroscopic superposition
$\longrightarrow$ very large entanglement involving microscopic 
degrees of freedom $\longrightarrow$
computational instability $\longrightarrow$ classical regime via rapid
cycle of information transitions and unitary evolution. 

As in the case of measurement, the environment is not needed for
``classicalization" in the sense that for a sufficiently large cat, it may
be true that ${\cal M}_{\rm cat} > {\cal M}_2$. However, interaction with the
environment is inevitable in most usual cases, and environmental decoherence 
helps explain localization of particles. 
As shown by Joos and Zeh
\cite{joo85}, decoherence due to scattering processes frequently causes this
effect by means of unitary destruction of
coherence between macroscopically separated positions.
 Consider the scattering of an
object in state $|\chi\rangle$ off a particle in state $\int_x\phi(x)|x\rangle d^3x$:
\be
\label{scatter}
\int_x d^3x\phi(x)|x\rangle|\chi\rangle \stackrel{t}{\longrightarrow} 
\int_x d^3x\phi(x)|x\rangle S_x|\chi\rangle,
\ee  
where the scattered state $S_x|\chi\rangle$ may be calculated using an
appropriate S-matrix. The reduced density operator for the scatterer changes as
$\phi(x)\phi^*(x^{\prime}) \longrightarrow \phi(x)\phi^*(x^{\prime})\langle
\chi|S^{\dag}_{x^{\prime}}S_x|\chi\rangle$. A single scattering like this cannot
quench the off-diagonal terms. Let $k, F, \sigma_{\rm eff}$ be, respectively,
the wave-number, flux and total cross section of the incoming particles. 
Letting $\Lambda \equiv k^2F\sigma_{\rm eff}$, one finds:
\be
\label{scaquench}
\rho(x,x^{\prime},t) = \rho(x,x^{\prime},0)\exp\{-\Lambda t(x - x^{\prime})^2\}
\ee
implying an exponentially rapid dying off of the off-diagonal terms \cite{joo85}.

Combining the damping in Eq. (\ref{scaquench})
due to decoherence with ``free" Schr\"odinger evolution,
we obtain:
\be
\label{d+f}
\frac{\partial\rho(x; x^{\prime}; t)}{\partial t} = \frac{i}{2m}
\left(\frac{\partial^2}{\partial x^2} - \frac{\partial^2}{\partial
x^{\prime^2}}\right)\rho -  \Lambda(x - x^{\prime})^2\rho
\ee
where, the first term in right hand side represents the free evolution with 
$H_0 = p^2/2m$, in coordinate representation, in the Lindblad equation Eq. (\ref{eqlind}).
One obtains the non-Hamiltonian part in Eq. (\ref{d+f}) as a special case of the 
Lindblad equation, setting
$\hat{L}^{\dag}_k = \hat{L}_k$. In this case, the Lindblad terms can be re-written as
$\hat{L}^2\rho + \rho \hat{L}^2 - 2\hat{L}\rho \hat{L} = 
[\hat{L},[\hat{L},\rho]]$. For $\hat{L}=\sqrt{2\Lambda} \hat{x}$,
one recovers Eq. (\ref{d+f}). 

In practice, the localizing tendency in Eq. (\ref{scaquench}) is to be balanced
against the natural spreading tendency of wavepackets. This suggests that there
is a finite scale, which we call the 
decoherence length-scale $\lambda_d$, representing 
the `equilibrium' size of wavepackets. Because coherence is more or less
preserved at length-scales smaller than
$\lambda_d$, phenomena confined to such small scales are effectively 
unaffected by information transitions, and behave quantum mechanically
even in a macroscopic body, provided $\lambda_d < \lambda_u$. Otherwise,
the relevant quantity is the minimum of the two length-scales. 
Microscopic objects such as electrons in atoms or photons in a conventional quantum
optical experiment will move through such structures according to the rules of
quantum mechanics, rather than classical mechanics, if the length
scale of the potentials they are responding to is smaller than 
${\rm min}\{\lambda_d,\lambda_u\}$, which scale can,
from this viewpoint, be regarded as the border
scale between microscopic and macroscopic phenomena.

It is worth noting that macroscopic matter is not necessarily classical. As is well
known, quantum behaviour has been noted in macroscopic systems manifesting coherent
matter states like Bose-Einstein condensates (BECs) \cite{ketterle} and in
Schr\"odinger cat states observed in superconducting quantum interference 
devices (SQUIDs) \cite{fri00}. In such systems, the microscopic degrees of 
freedom are frozen, as a result of which ${\cal M} \ll {\cal M}_{2}$.
Indeed, whereas $n$-partite entanglement requires exponentially 
more memory support than for
$n$ particles in separable states, BECs, superconductors
or some other such coherent state matter, require {\em less}. 
From this standpoint, condensation is the opposite of entanglement.
A condensate has almost zero entropy in that all particles collectively 
occupy the same single many-particle ground state. 
For example, the Schr\"odinger cat states observed in SQUID superconductors
are described by macroscopic clockwise-counterclockwise
oscillations of the flux through a loop, a collective coordinate representing
the motion of $\sim 10^9$ Cooper pairs acting in tandem. Since the experimental
temperature is 1000 times smaller than the superconducting energy gap, almost
most microscopic degress of freedom are frozen out and only the collective flux
coordinate retains any dynamical relevance \cite{fri00}. 

Let us consider a system of $n$ elementary 2-state
objects restricted to volume ${\cal V}$. As seen
earlier, state information for the system if the particles are non-interacting,
and thus their states remain separable, is ${\cal M}_{\rm separable} 
= n10^{182}\mu$.
In the case of ordinary matter, where interaction causes entanglement,
${\cal M}_{\rm usual} = 10^{182n}\mu$. 
Suppose the system is a condensate. As its
coherent matter involves elementary objects (atoms or Cooper pairs of electrons) 
`coalescing' into and behaving like a single, coherent, non-interacting whole,
the required memory is ${\cal M}_{\rm coherent} = 
10^{182} \mu \ll {\cal M}_2$. 
Considering that we estimated ${\cal M}_2$ to be sufficiently large as to support
even the separable evolution of all particles in the universe, it follows that coherent
matter will not suffer information transitions at all, and can be treated as entirely
quantum mechanical.  

\section{Conclusions}

The information processing involved in the evolution of physical systems, viewed as 
computations performed by nature,
were quantified using the physics of information processing. 
The role of entanglement in exponentially augmenting its requirement
was pointed out. In
particular, in a system with $n$ entangled subsystems of average energy $E$
and identically of dimension $D$,
computational speed requirement goes as $\sim 2^{\mu/2}D^nnE$ op/s, whilst
memory allocation requirement goes as $\sim D^n\mu$. However, most of this
is physically inaccessible, essentially because of Heisenberg uncertainty,
except when suitably designed
interference manifests it, as in a Young's double-slit experiment or Shor's
prime factorization algorithm \cite{shor}. 
Why is only a fraction of the underlying state information accessible for classical 
information processing? From the standpoint of the present model, 
it might well have more to do with 
physical laws rather than with the information theoretic structure of Hilbert space.
However, interestingly, such a probabilistic limit on accessible
information has been argued as indicative of the informational nature of quantum
mechanics \cite{zei01,mohm02}. 

Is nature a computer? It is certainly not a digital computer running Linux or Windows. 
That information processing and 
computations actually underlie, rather than merely offer an interprettation of, 
the existence and evolution of physical reality is a controversial yet rich
scientific paradigm that has been explored to various degrees in other contexts
\cite{cah96,zei01,wol02,mohm02}. In our model, we find that the classicality of
measurement outcomes and of usual macroscopic systems
reflects an underlying information processing threshold, related to the finite
fine-graining of Hilbert space, which is highly suggestive of an information
theoretic origin of physical reality.  In this viewpoint, nature 
can arguably be thought of as an abstract operating system running a fundamental
{\em unitary} program, that is continually interrupted by
entanglement-engendered memory-overflows at macroscopic scales. 

The dynamics of the information transitions depends on the computational architecture of
this `operating system' underlying 
physical systems, that may not
appear in the arena of physical dynamics but instead at the deeper sub-quantum level. 
Here it is of interest to study whether 
the free parameter of the computational model, namely $\mu$,
emerges as a basic information processing feature of nature, that
can be linked in intresting ways to other aspects of physics, in particular
quantum gravity and quantum cosmology.
The computational model complements decoherence in quantum measurements, by
turning the improper mixture of decoherentially localized particles
of the subsystem of interest into
a proper mixture. It thereby helps {\em complete} the measurement act and
offers a resolution to the measurement problem.
It does so adding arguably minimal new physics in almost all experimentally
accessible regimes.

The model implies that classicality at the macroscopic level arises because
ever-increasing entanglement can `overwhelm' nature's computational capacity.
In view of Eq. (\ref{gammao}), it roughly reflects the fact that $\mu \ll N_p 
\approx 10^{120}$, where
$N_p$ is the number of degrees of freedom in the universe. 
In a universe where $\mu \gg N_p$, quantum mechanics would possess enough
mnemonic and computational resource to support the entangled evolution of all
matter and energy. Information transitions, and hence
the classicality of measurement outcomes and macroscopic matter, would be absent. 
Macroscopic phenomena will regularly be quantum mechanical.
Decoherence will still causes localization of particles, but superpositions of
such decohered/localized particles will persist. So, why is $\mu$
relatively so small? The question can be paraphrased conversely:
why is the universe relatively so large? This is analogous to a situation
in astronomy: that the large scale evolution of the universe is slow can be viewed as
due to the relative slowness of light's speed, or as due to the immense size of the
universe.  One can also ask if there
is some sort of Dirac-like large number hypothesis \cite{llo02} 
that can connect $\mu$ and $N_p$.

The idea of states of physical systems and their evolution deriving from an
information processing substratum raises
deep philosophical questions: 
What is the basic hardware on which the `operating
system' called nature is mounted? Can one go farther and, as suggested in Section
\ref{compu}, regard physical objects 
themselves as information theoretic constructs? What sort of tests would be needed
to elevate such a hypothesis beyond the status of a mere interpretation
to an ontology 
(just as macroscopic classicality is said to indicate the computational
nature of physical evolution of systems)? And
what does this imply for and how does it relate to
quantum gravity, a program that, inspite of the recent
spectacular successes in string and M theories \cite{free} and loop quantum gravity
\cite{loop}, proves to be a very difficult and elusive goal?
It is hoped that further
developments in the relation between quantum physics and computation will shed
light on these, and related, questions. 

What are there decisive tests of the model? The very classicality of measurement
outcomes and macroscopic matter is, in our opinion, a relevant evidence, 
in view of the Bassi-Ghirardi theorem \cite{bas00}. However, the evidence is not
clinching, because collapse theories and Many-worlds interpretations also predict the
same. Other than this, evidence from mesoscopic systems, specifically
experiments of the types described in Refs. \cite{c70,alt02}, will be important. This
is because the predictions of the computational model are indistinguishable from
quantum mechanics for microscopic systems and open macroscopic systems. 
Now, like collapse models, 
the computational model implies that even closed macroscopic systems exist in mixed 
state and show irreversible behaviour. In practice, it is rather difficult to test 
this, because there are rarely truely isolated macroscopic systems, other than
condensates and superfluids, which, as noted earlier, present qualitatively
different situations. 
Even interstellar dust grains are constantly scattering cosmic background
radiation and are thus open \cite{joo00}. 

To be able to resolve the matter, careful study of isolated, mesoscopic
systems is needed. This is no easy task experimentally, not only from the
viewpoint of noise-control, but also because the information transitional
effects would have to be distinguished from quantum chaotic and decoherence
effects. But such studies alone can shed more
light on the onset of decoherence and the emergence of computational instability
and also enable us to fix $\mu$, the free
parameter for the computational model, and thence ${\cal M}_2$ and $\lambda_u$.

I am thankful to Messrs. Kaushik Mitra and Sudhir Rao for discussions and
pointing out important references. I thank Prof. J. Pasupathy for useful
discussions. This work, begun at the Center for Theoretical Studies,
Indian Institute of Science, was partially supported by the DRDO510 project no. 
01PS:00356. 


\begin{thebibliography}{100}
\bibitem{multiverse} For example, if the initial state of $S$ in Eq. (\ref{envent}) is
$(1/\sqrt{2})((1/\sqrt{23}|0\rangle + \sqrt{22/23}|1\rangle)$ in the measurement
basis, it would suggest
that a measurement in the $\{|0\rangle,|1\rangle\}$ basis, in order to conform to
the Born probability law, must cause the universe to branch out
into 23 versions, exactly 1 of which goes into the $|0\rangle$ state. And if the
coefficients $a_i$ are irrational, then each measurement would seem to cause the
universe to branch out into infinite versions.
\bibitem{hp} According to the
holographic principle \cite{ng}, the maximum information that can be
registered in a system is $A/l_P^2$, where $A$ is the system's area. This would
suggest ${\cal M} = \mu (V/V_P)^{2/3}$. This may be relevant in the quantum
gravity regime, but for our purposes, we will quanitfy ${\cal M}$ as the volume
Kolmogorov-Chaitin complexity.
\bibitem{codrat} 
Actually, the required clock rate should be $\omega_{\chi} ({\cal D}^{\prime}{\cal P}/{\cal
M})$ where $x = ((1/r) - 1)$ and ${\cal P}$ is the number of elements
in the parity check matrix \cite{coderate}, given by 
${\cal P} = \mu^2n^2/r$. For our estimate we ignore ${\cal P}$, which is
polynomial in $n$. Also note that we are invoking a simple classical error correction
procedure,
rather than quantum error correction \cite{css}, as the correction is supposed to
be implemented on the state information rather than on quantum states.
\bibitem{sf} Segmentation fault occurs when there is invalid pointer de-referencing.
In our case, the pointer is invalid because it is trying to access a piece of
unallocated (or de-allocated) memory. In conventional computing, it can be invalid
also because it is a NULL-pointer or an un-initialized pointer.
\bibitem{cd} The core is the file that contains a snapshot of the memory registers at the 
time when a program crashes. It is 'dumped' into a compiler-specified directory
when a program crashes.
\bibitem{natbar} In fact, computational stability is only necessary but not sufficient to
manifest coherent wavelike behaviour.
In practice, it will also be necessary to take into consideration various energy or
wavelength scales of the system (rotational, vibrational, etc.). % as being
% too large or too small to affect the manifestation of coherent wavelike behaviour. 
The presence of such multiple internal degrees of freedom of comparable energy 
can lead to broad-band
wavelength scales that can potentially wash out interference patterns.
\end{thebibliography}
\end{document}